\documentclass[reprint, amsmath,amssymb, nofootinbib, aps, prd]{revtex4-2}

\usepackage{mathtools}
\usepackage{graphicx}
\usepackage[dvipsnames]{xcolor}
\usepackage[colorlinks,allcolors=Blue]{hyperref}

\newcommand{\mnras}{Mon. Not. Roy. Astron. Soc.}
\newcommand{\aap}{Astronomy \& Astrophysics}

\begin{document}

\title{A new scale in the quasi-static limit of Aether Scalar Tensor Theory}

\author{Tobias Mistele}%
\email{tobias.mistele@case.edu}
\affiliation{Department of Astronomy, Case Western Reserve University\\
10900 Euclid Avenue,
44106 Cleveland, Ohio, USA}

\begin{abstract}
One of the aims of Aether Scalar Tensor Theory (AeST) is to reproduce the successes of Modified Newtonian Dynamics (MOND) on galactic scales.
Indeed, the quasi-static limit of AeST achieves precisely this, assuming that the vector field $\vec{A}$ vanishes and that the so-called ghost condensate can be neglected.
The effects of the ghost condensate were investigated in detail in previous studies.
Here, we focus on the assumption of a vanishing vector field.
We argue that this assumption is not always justified and show how to correctly take into account the vector field, finding that the quasi-static limit depends on a model parameter $m_\times$.
In the limit $m_\times \to 0$, one recovers the quasi-static limit with a vanishing vector field.
In particular, one finds a two-field version of MOND.
In the opposite limit, $m_\times \to \infty$, one finds a single-field version of MOND.
We show that, in practice, much of the phenomenology of the quasi-static limit depends only very little on the value of $m_\times$.
Still, for some observational tests, such as those involving wide binaries, $m_\times$ has percent-level effects that may be important.
\end{abstract}

\maketitle

\section{Introduction}
\label{sec:introduction}

Aether Scalar Tensor Theory \citep[AeST,][]{Skordis2020, Skordis2021} is a fully-relativistic model that reproduces much of the successes of both  Modified Newtonian Dynamics \citep[MOND,][]{Milgrom1983a, Milgrom1983b, Milgrom1983c} on galactic scales and of $\Lambda$CDM on cosmological scales.
It also satisfies observational constraints from gravitational waves that require tensor modes to propagate at the speed of light \citep{Sanders2018, Boran2018}.

One success of MOND that AeST aims to reproduce is explaining the observed Radial Acceleration Relation \citep[RAR,][]{Lelli2017b, Brouwer2021, Mistele2023d}.
The RAR shows a one-to-one correspondence between the Newtonian acceleration due to baryons, $a_{b}$, and the actually observed acceleration, $a_{\mathrm{obs}}$.
In particular, for accelerations $a_{b}$ much larger than the acceleration scale $a_0 \approx 10^{-10}\,\mathrm{m/s}^2$, the total acceleration $a_{\mathrm{obs}}$ is just the Newtonian baryonic acceleration $a_{b}$.
At small accelerations, $a_{b} \ll a_0$, the total acceleration $a_{\mathrm{obs}}$ is instead given by $ a_{\mathrm{obs}} \approx \sqrt{a_0 \, a_{b}}$.

That AeST can indeed reproduce MOND phenomenology was shown in \cite{Skordis2020} under the assumption that the vector field $\vec{A}$ (see below) vanishes.
Their argument establishes that AeST reduces to a specific two-field version of MOND as long as the mass of the so-called ghost condensate is negligible.
The scale where the ghost condensate becomes important is set by a model parameter that we call $m$ below.
That is, $m$ controls whether AeST reproduces MOND.
This scale and the effects of the ghost condensate are investigated in detail in previous studies \citep{Skordis2020, Mistele2023,Verwayen2023}.

Here, we focus on the assumption of a vanishing vector field $\vec{A}$.
In general, setting $\vec{A}$ to zero is inconsistent \citep{Mistele2023}.
Nevertheless, one might expect it to be a good approximation in many situations of practical interest.
Indeed, neglecting $\vec{A}$ amounts to neglecting a curl term and, in the context of MOND, neglecting curl terms was often found to be a good approximation (see, e.g., \cite{Bekenstein1984}).

In some cases, however, curl terms can be phenomenologically important.
In addition, as we will see, neglecting the vector field in AeST means neglecting a term whose prefactor can be arbitrarily large, depending on the choice of a model parameter.
Thus, one might worry that the vector field cannot be neglected in cases where this prefactor is large.
Here, we demonstrate that, even in these cases, the ability of AeST to reproduce MOND is not impaired.
That is, AeST still reproduces MOND as long as the ghost condensate is negligible.
More generally, we show how to correctly take the vector field $\vec{A}$ into account and discuss phenomenological implications.

In particular, we will find that the quasi-static limit of AeST depends on an additional mass parameter we call $m_\times$.
This mass parameter $m_\times$ should be distinguished from the mass parameter $m$ that controls the ghost condensate.
We will see that, in contrast to $m$, the new mass parameter $m_\times$ is not related to the ghost condensate, has no effect in spherical symmetry, and does not affect the ability of AeST to reproduce MOND.

In the following, we employ units with $c = \hbar = 1$.

\section{The model}
\label{sec:model}

In the quasi-static weak-field limit and with only non-relativistic matter sources, the action $S$ of AeST can be written in terms of the Newtonian gravitational potential $\Phi$, a scalar field $\varphi$, and a vector field $\vec{A}$ \citep{Skordis2020,Mistele2023},\footnote{
  The derivation follows \cite{Skordis2020}.
  We start with a Newtonian gauge for the metric and use the assumption of a weak-field limit with no time-dependence (except $\dot{\varphi} = \mathrm{const}$, see below).
  Then, the $ij$ components of the Einstein equations imply that the metric depends only on a single potential $\Phi$, which we substitute back into the action.
  This last step works because the fields $\varphi$ and $\vec{A}$ do not couple to the metric components $g_{ij}$ in this quasi-static weak-field limit.
}
\begin{widetext}
\begin{multline}
 \label{eq:actionwithA}
 8 \pi \hat{G} \, S =
 - \int d^4x \left\{(\vec{\nabla} \Phi)^2 - 2 \vec{\nabla} \Phi \, (\vec{\nabla} \varphi + Q_0 \vec{A})  \right.
 + (\vec{\nabla} \varphi + Q_0 \vec{A})^2
 + \mathcal{J}\left((\vec{\nabla} \varphi + Q_0 \vec{A})^2\right)
 \\
- m^2 \left(\frac{\dot{\varphi}}{Q_0} - \Phi\right)^2
+ \frac{2K_B}{2-K_B} \vec{\nabla}_{[i} \vec{A}_{j]}  \vec{\nabla}^{[i} \vec{A}^{j]}
+ 8 \pi \hat{G} \, \Phi \rho_b
\bigg\}
\,.
\end{multline}
\end{widetext}
Here, $\rho_b$ denotes the matter density.
We use the notation $\rho_b$ because only the baryonic matter is relevant in the cases we discuss explicitly below.
The function $\mathcal{J}$ determines how the model interpolates between the Newtonian regime (for accelerations larger than $a_0$) and the MOND regime (for accelerations smaller than $a_0$).
That is, $\mathcal{J}$ determines what is called the interpolation function in MOND \citep{Famaey2012}.
Further, $\hat{G}$, $Q_0$, $m$, and $K_B$ are constants with mass dimension $-2$, $1$, $1$, and $0$, respectively.
In the following, we sometimes use the notation
\begin{align}
 \label{eq:Udef}
 \vec{U} \equiv \vec{\nabla} \varphi + Q_0 \vec{A} \,.
\end{align}

We do not set $\dot{\varphi}$ to zero because any $\dot{\varphi} = \mathrm{const}$ also allows for static solutions.
Indeed, the AeST model contains a so-called ghost condensate and $\dot{\varphi}$ represents the chemical potential of this condensate.
This is explained in more detail in \cite{Mistele2023}.
The term multiplied by $m^2$ is related to the condensate's energy density.

The constant $\hat{G}$ is related to the Newtonian gravitational constant $G_N$ as
\begin{align}
 \label{eq:fGdef}
 G_N \equiv \hat{G} \, f_G^{-1} \,,
\end{align}
with constant $f_G$.
Here, $G_N$ is what enters the total acceleration felt by matter in the Newtonian regime, i.e. at accelerations larger than $a_0$.
That is, in spherical symmetry and in the Newtonian regime, the total acceleration has the form $G_N M/r^2$ with the total mass $M$ and the spherical radius $r$.

The action Eq.~\eqref{eq:actionwithA} depends on the four model parameters $\hat{G}$, $Q_0$, $m$, and $K_B$, but only three independent combinations of these are physical.
Below we will mostly use the following three independent combinations,
\begin{align}
f_G, \, m, \,  m_\times \equiv Q_0 \sqrt{\frac{2-K_B}{2 K_B}}.
\end{align}
To see that this is possible, rescale $\vec{A} \to \vec{A}/Q_0$ (as in the definition of $\vec{U}$ from Eq.~\eqref{eq:Udef}).
Then the prefactor of the $\vec{A}$ kinetic term becomes $1/m_\times^2$ and the only remaining appearance of $Q_0$ is in the $m^2(\dot{\varphi}/Q_0 - \Phi)^2$ term.
In the quasi-static limit that we consider, $\dot{\varphi}/Q_0$ can be absorbed into $\Phi$.
Thus, only $f_G$, $m$, and $m_\times$ are physical.

The function $\mathcal{J}$ must satisfy some requirements that ensure that there is a Newtonian regime at large accelerations and a MOND regime at small accelerations.
In particular, the interpolation function $\tilde{\mu}(|\vec{U}|/a_0) \equiv \mathcal{J}'(\vec{U}^2)$ must have the limits \citep{Skordis2020, Famaey2012},
\begin{subequations}
\label{eq:mutildelimits}
\begin{align}
 \left.\tilde{\mu}\left(\frac{|\vec{U}|}{a_0}\right)\right|_{\frac{|\vec{U}|}{a_0} \to 0} &= f_G \, \frac{|\vec{U}|}{a_0} \,, \\
 \left.\tilde{\mu}\left(\frac{|\vec{U}|}{a_0}\right)\right|_{\frac{|\vec{U}|}{a_0} \to \infty} &= \frac{f_G}{1-f_G} \,.
\end{align}
\end{subequations}
In the following, we adopt $a_0 = 1.2 \cdot 10^{-10}\,\mathrm{m/s}^2$ \citep{Lelli2017b}.

In AeST, ordinary matter is minimally coupled to the metric $g_{\mu \nu}$ in the standard way.
The metric has the same form as in the Newtonian limit of General Relativity with the potential $\Phi$.
There is no additional coupling of matter to the fields $\varphi$ and $\vec{U}$.
Thus, the total acceleration felt by matter is
\begin{align}
\vec{a}_{\mathrm{tot}} \equiv - \vec{\nabla} \Phi \,.
\end{align}
Like matter, light is also minimally coupled to the metric $g_{\mu \nu}$ in the standard way (although this is not shown explicitly in Eq.~\eqref{eq:actionwithA}).
Thus, lensing works as in General Relativity, just with the potential $\Phi$ calculated from different field equations \citep{Skordis2021, Mistele2023}.

\section{Why one may be cautious about leaving out the vector field}
\label{sec:novector}

To set the stage, we first assume a vanishing vector field $\vec{A}$ and show that this assumption is, in general, inconsistent.
Nevertheless, as mentioned above, one may expect that neglecting $\vec{A}$ is a good approximation in many situations of practical interest.
Here, we discuss why this is the case and why, in some cases, one may still be cautious about neglecting $\vec{A}$.

With $\vec{A}$ set to zero, the equations of motion for the remaining fields $\varphi$ and $\Phi$ read \citep{Skordis2020},
\begin{subequations}
\label{eq:eomA0}
\begin{align}
\Delta \hat{\Phi} &= f_G \cdot 4 \pi G_N \left(\rho_b + \rho_c\right) \,, \\
\vec{\nabla} \left( \tilde{\mu}\left(\frac{|\vec{\nabla} \varphi|}{a_0}\right) \vec{\nabla} \varphi \right) &=  f_G \cdot 4 \pi G_N \left(\rho_b + \rho_c \right) \,,
\end{align}
\end{subequations}
where we defined
\begin{align}
 \hat{\Phi} = \Phi - \varphi \,,
\end{align}
and where $\rho_c$ is the density of the ghost condensate,
\begin{align}
 \label{eq:rhoc}
 \rho_c = \frac{m^2}{4 \pi G_N f_G} \left(\frac{\dot{\varphi}}{Q_0} -\hat{\Phi} - \varphi\right) \,.
\end{align}
The total acceleration $\vec{a}_{\mathrm{tot}} = - \vec{\nabla} \Phi$ can be written as the sum of the accelerations $\vec{a}_\varphi = -\vec{\nabla} \varphi$ and $\vec{a}_{\hat{\Phi}} = - \vec{\nabla} \hat{\Phi}$ due to the fields $\varphi$ and $\hat{\Phi}$, respectively,
\begin{align}
 \vec{a}_{\mathrm{tot}} = \vec{a}_{\varphi} + \vec{a}_{\hat{\Phi}} \,.
\end{align}

These equations give MOND-like behavior whenever the effects of the condensate density $\rho_c$ are negligible.
To see this, consider a spherically symmetric system and consider first the large-acceleration regime $a_b \gg a_0$.
Then, using the limits of $\tilde{\mu}$ from Eq.~\eqref{eq:mutildelimits}, we find
\begin{align}
 a_\varphi = (1-f_G) a_b \,, \quad a_{\hat{\Phi}} = f_G \, a_b \,,
\end{align}
which gives
\begin{align}
 a_{\mathrm{tot}} = a_b \,,
\end{align}
which is exactly what MOND requires in this large-acceleration regime.
Similarly, in the small-acceleration regime, $a_b \ll a_0$, we find
\begin{align}
 a_{\mathrm{tot}} \approx a_\varphi \approx \sqrt{a_0 a_b} \,,
\end{align}
which again matches MOND.
Thus, AeST seems to have achieved its goal of reproducing MOND-like behavior.
In particular, AeST seems to reproduce a multifield version of MOND whenever the ghost condensate can be neglected \citep{Famaey2012}.

The problem is that setting $\vec{A}$ to zero is not always allowed.
The reason is that the equations of motion of $\varphi$ and $\Phi$, i.e. Eq.~\eqref{eq:eomA0}, are not the only equations that must be satisfied.
The equation of motion of $\vec{A}$ must be satisfied as well.
And this $\vec{A}$ equation of motion does not in general have $\vec{A} = 0$ as a solution.
Concretely, the $\vec{A}$ equation of motion reads
\begin{multline}
\label{eq:Aeom}
\vec{\nabla} \Phi - \frac1{2 m_\times^2} \vec{\nabla} \times \vec{\nabla} \times (Q_0 \vec{A}) = \\
(\vec{\nabla} \varphi + Q_0 \vec{A}) \left(1 + \tilde{\mu}\left(\frac{|\vec{\nabla} \varphi + Q_0 \vec{A}|}{a_0}\right)\right) \,.
\end{multline}
If it is allowed to set $\vec{A}$ to zero, this equation must be satisfied with $\vec{A} = 0$,
\begin{align}
 \label{eq:AeqA0}
\vec{\nabla} \Phi = \vec{\nabla} \varphi \, \left(1 + \tilde{\mu}\left(\frac{|\vec{\nabla} \varphi|}{a_0} \right)\right) \,.
\end{align}
We can algebraically solve this equation for $\vec{\nabla} \varphi$ and then take the curl to find
\begin{align}
 \label{eq:PhiA0cond}
 \vec{\nabla} |\vec{\nabla} \Phi| \times \vec{\nabla} \Phi = 0 \,.
\end{align}
This condition is not fulfilled, except in some special cases like spherical symmetry \citep{Brada1995}.\footnote{
  There is an alternative way to see that $\vec{A} = 0$ is not, in general, allowed.
  On the one hand, by subtracting the $\varphi$ and $\Phi$ equations of motion from Eq.~\eqref{eq:eomA0} from each other, we find $\vec{\nabla} \left( \tilde{\mu} \cdot \vec{\nabla} \varphi - \vec{\nabla} \hat{\Phi}\right) = 0$ where we left out the argument of $\tilde{\mu}$ for brevity.
  This implies that $\tilde{\mu} \cdot \vec{\nabla} \varphi - \vec{\nabla} \hat{\Phi} = \vec{\nabla} \times \vec{h}$ for some vector field $\vec{h}$.
  Here, $\vec{\nabla} \times \vec{h}$ is nonzero except in some special cases such as spherical symmetry \citep{Bekenstein1984}.
  On the other hand, with $\vec{A} = 0$, the $\vec{A}$ equation of motion Eq.~\eqref{eq:Aeom} can be written as $\tilde{\mu} \cdot \vec{\nabla} \varphi - \vec{\nabla} \hat{\Phi} = 0$, which implies $\vec{\nabla} \times \vec{h} = 0$.
  This is a contradiction (except in special cases such as spherical symmetry).
  This contradiction is resolved by allowing (the curl part of) $\vec{A}$ to be non-zero.
}
This is important because for many systems of astrophysical interest -- such as disk galaxies -- one cannot assume spherical symmetry or another of the special cases where Eq.~\eqref{eq:PhiA0cond} holds.

Thus, setting $\vec{A}$ to zero seems to be inconsistent except in special cases such as spherical symmetry.
As we will see below in Sec.~\ref{sec:twofield}, an exception  is the limit $m_\times \to 0$.
In that limit, a small $\mathcal{O}(m_\times^2)$ curl term in $\vec{A}$ allows to both avoid Eq.~\eqref{eq:PhiA0cond} and use the equations of motion Eq.~\eqref{eq:eomA0} that were derived assuming $\vec{A} = 0$.
In that sense, setting $\vec{A}$ to zero is allowed in the $m_\times \to 0$ limit, even outside special cases such as spherical symmetry.
The inconsistency remains, however, when neither the limit $m_\times \to 0$ nor a special case such as spherical symmetry is pertinent. 

In fact, this inconsistency arises already when setting only the curl part of $\vec{A}$ to zero while still allowing a non-zero gradient part.
That is, the same inconsistency arises when enforcing that $\vec{A}$ is of the form $\vec{A} = \vec{\nabla} \alpha_A$ for some $\alpha_A$.
Indeed, a symmetry of the quasi-static limit of AeST allows absorbing such a gradient term into $\vec{\nabla} \varphi$ so that a vector field $\vec{A} = \vec{\nabla} \alpha_A$ is equivalent to $\vec{A} = 0$ \cite{Skordis2021}.

To see this more explicitly, consider a Helmholtz decomposition of $\vec{A}$,
\begin{align}
 \vec{A} = \vec{\nabla} \alpha_A + \vec{\nabla} \times \vec{\beta}_A \,,
\end{align}
where $\vec{\beta}_A$ parametrizes the curl part of $\vec{A}$.
Using this decomposition in Eq.~\eqref{eq:Aeom} and setting the curl part to zero, $\vec{\beta}_A = 0$, we obtain an equation of the same form as Eq.~\eqref{eq:AeqA0}, just with $\varphi$ replaced by $\chi \equiv \varphi + Q_0 \alpha_A$,
\begin{align}
\vec{\nabla} \Phi = \vec{\nabla} \chi \, \left(1 + \tilde{\mu}\left(\frac{|\vec{\nabla} \chi|}{a_0} \right)\right) \,.
\end{align}
The problematic equation Eq.~\eqref{eq:PhiA0cond} and, therefore, the inconsistency then follow as before, i.e. by algebraically solving for $\vec{\nabla} \chi$ and then taking the curl.

In the context of MOND, neglecting curl terms is often a good approximation (see, e.g., \cite{Bekenstein1984}).
Since what is problematic about setting $\vec{A}$ to zero is really setting its curl part $\vec{\beta}_A$ to zero, one might expect a similar result here.
That is, in practice, neglecting (the curl part of) $\vec{A}$ may still be a good approximation.

However, there is also reason to be cautious.
In AeST, neglecting the curl part of $\vec{A}$ implies neglecting the following term in Eq.~\eqref{eq:Aeom},
\begin{align}
 - \frac{1}{2 m_\times^2} \vec{\nabla} \times \vec{\nabla} \times (Q_0 \vec{A}) \,.
\end{align}
The prefactor of this term can be arbitrarily large, depending on the choice of the model parameter $m_\times$.
Thus, at least when this prefactor is large, one may worry that neglecting $\vec{A}$ is a bad approximation.
Below, we show that this particular case is not actually problematic and, more generally, show how to correctly take the vector field into account.

\section{Taking the vector field into account}
\label{sec:vector}

We now show how to correctly take the vector field $\vec{A}$ into account and establish that this does not affect the ability of AeST to reproduce MOND-like behavior.
That is, we show that AeST still reproduces MOND whenever the ghost condensate can be neglected.
We discuss the phenomenological implications in Sec.~\ref{sec:mw} and Sec.~\ref{sec:wb}.

We first write the action in terms of $\vec{U}$ (see Eq.~\eqref{eq:Udef}),
\begin{multline}
 \label{eq:actionwithU}
 S = - \int d^4x \left\{\frac{1}{8 \pi \hat{G}} \left[ (\vec{\nabla} \Phi)^2 - 2 \vec{\nabla} \Phi \, \vec{U} + \vec{U}^2 + \mathcal{J}\left(\vec{U}^2\right)\right.\right.\\
\left.\left. - m^2 \left(\frac{\dot{\varphi}}{Q_0} - \Phi\right)^2  +
\frac{1}{m_\times^2} \vec{\nabla}_{[i} \vec{U}_{j]}  \vec{\nabla}^{[i} \vec{U}^{j]}  \right] + \Phi \rho_b \right\} \,,
\end{multline}
where we used $Q_0^2 \, \vec{\nabla}_{[i} \vec{A}_{j]} = \vec{\nabla}_{[i} \vec{U}_{j]}$ which follows from antisymmetry, and we defined the mass scale
\begin{align}
 m_\times \equiv \sqrt{\frac{2 - K_B}{2 K_B}} Q_0 \,.
\end{align}
The equation of motion of $\varphi$ is then of the form $\partial_t(\dot\varphi - Q_0 \Phi) = 0$ which is trivially satisfied in the quasi-static limit (keeping in mind that the only time-dependence we allow is $\dot{\varphi} = \mathrm{const}$).\footnote{
  This assumes a change of variables from $(\vec{A}, \varphi)$ to $(\vec{U}, \varphi)$.
  In terms of $(\vec{A}, \varphi)$, the $\varphi$ equation of motion still reduces to $\partial_t(\dot{\varphi} - Q_0 \Phi) = 0$ when combined with the divergence of the $\vec{A}$ equation of motion.
}
This leaves the equations of motion of $\Phi$ and $\vec{U}$.
They read
\begin{subequations}
\label{eq:eom}
\begin{align}
 \label{eq:eom:Phi}
 \Delta \Phi - \vec{\nabla} \vec{U} &= 4 \pi G_N f_G (\rho_b + \rho_c) \,, \\
 \label{eq:eom:U}
 \vec{\nabla} \Phi - \frac{1}{2 m_\times^2} \vec{\nabla} \times \vec{\nabla} \times \vec{U} &= \vec{U} \left(1 + \tilde{\mu}\left(\frac{|\vec{U}|}{a_0}\right) \right) \,.
\end{align}
\end{subequations}
These are the equations that must be solved in the quasi-static limit of AeST.
They replace Eq.~\eqref{eq:eomA0} when one correctly takes the vector field into account.
As we will see, they are equivalent to the equations Eq.~\eqref{eq:eomA0} in some special cases such as in the limit $m_\times \to 0$ (Sec.~\ref{sec:twofield}).
Another, independent, special case where Eq.~\eqref{eq:eom} and Eq.~\eqref{eq:eomA0} are equivalent is spherical symmetry (Sec.~\ref{sec:algebraic}).

To see that the equations Eq.~\eqref{eq:eom} can still describe MOND-like behavior, we will first discuss the limits of large and small $m_\times$.
We will find that both describe a version of MOND.
For $m_\times \to 0$, one recovers the two-field version of MOND from Eq.~\eqref{eq:eomA0}.
That is, setting $\vec{A}$ to zero is justified in this case (but not in general).
In the opposite limit, $m_\times \to \infty$, one recovers a single-field version of MOND.
Plausibly, other values of $m_\times$ interpolate between these single-field and two-field limits.

Two comments are in order here.
First, by recovering a version of MOND we mean recovering equations that produce MOND-like behavior to the same extent as the original equations Eq.~\eqref{eq:eomA0} derived in Ref.~\cite{Skordis2020} assuming a vanishing vector field.
That is, both the single-field limit and the two-field limit we discuss below reproduce MOND-like behavior as long as the effects of the ghost condensate are negligible.

Second, our interest here is not in infinitely large or infinitely small values of $m_\times$.
Rather, our interest is in keeping $m_\times$ finite and applying the model to systems with characteristic length scales $l$ that are small or large relative to $1/m_\times$, i.e. $m_\times l \gg 1$ or $m_\times l \ll 1$.
That $m_\times l$ is the relevant dimensionless quantity can be seen from Eq.~\eqref{eq:eom:U} by roughly estimating spatial derivatives to be of order $1/l$ (see also Sec.~\ref{sec:scale})
In the following, we obtain the leading-order behavior for these cases by considering the formal limits $m_\times \to \infty$ and $m_\times \to 0$.
But keep in mind that these limits are only a formal tool.
What we are really interested in here are the cases $m_\times l \gg 1$ and $m_\times l \ll 1$ with finite $l$ and finite $m_\times$.\footnote{
  This distinction is important for the following consideration.
  From the action Eq.~\eqref{eq:actionwithU} one sees that, for $m_\times \to \infty$, all spatial derivatives of the vector field $\vec{U}$ vanish.
  Indeed, in the full weak-field limit of AeST \cite{Skordis2021}, all derivatives of $\vec{U}$, including time derivatives, vanish for $m_\times \to \infty$.
  Thus, there may be a strong-coupling problem in the limit $m_\times \to \infty$.
  However, this is not a concern for us.
  The reason is that we only consider finite $m_\times$ and use the limit $m_\times \to \infty$ only as a formal tool within the quasi-static limit to obtain the leading-order behavior of systems whose characteristic length scale $l$ satisfies $m_\times l \gg 1$.
}

\subsection{The single-field limit $m_\times \to \infty$}
\label{sec:singlefield}

The complicated part of the equations of motion Eq.~\eqref{eq:eom} is the double-curl term proportional to $1/m_\times^2$ in the $\vec{U}$ equation of motion.
Luckily, for $m_\times \to \infty$, this term vanishes and we can algebraically solve for $\vec{U}$.
We choose to write the result in the following form,
\begin{align}
 \label{eq:mudef}
 \vec{U} = \vec{\nabla} \Phi \left(1 - f_G \, \mu\left(\frac{|\vec{\nabla} \Phi|}{a_0}\right) \right) \,,
\end{align}
which implicitly defines the function $\mu$ in terms of the function $\tilde{\mu}$.
From the limiting behavior of $\tilde{\mu}$, Eq.~\eqref{eq:mutildelimits}, we can infer the limiting behavior of $\mu$, namely
\begin{subequations}
 \label{eq:muimits}
\begin{align}
 \left.\mu\left(\frac{|\vec{\nabla} \Phi|}{a_0}\right)\right|_{\frac{|\vec{\nabla} \Phi|}{a_0} \to 0} &= \frac{|\vec{\nabla} \Phi|}{a_0} \,, \\
 \left.\mu\left(\frac{|\vec{\nabla} \Phi|}{a_0}\right)\right|_{\frac{|\vec{\nabla} \Phi|}{a_0} \to \infty} &= 1 \,.
\end{align}
\end{subequations}
This is the correct behavior for a single-field MOND interpolation function \citep{Famaey2012}.

Indeed, plugging this solution for $\vec{U}$ into the $\Phi$ equation of motion Eq.~\eqref{eq:eom:Phi} gives
\begin{align}
 \label{eq:eomsingle}
 \vec{\nabla} \left(\mu\left(\frac{|\vec{\nabla} \Phi|}{a_0}\right) \vec{\nabla} \Phi \right) = 4 \pi G_N (\rho_b + \rho_c) \,,
\end{align}
which is a standard single-field version of MOND \citep{Bekenstein1984, Famaey2012} up to the ghost condensate density $\rho_c$.
Without the ghost condensate, this is known as aquadratic Lagrangian theory (AQUAL).

This shows that the $m_\times \to \infty$ limit of AeST reproduces MOND-like behavior as long as the effects of the ghost condensate density are negligible.
Thus, in that sense, AeST reproduces a single-field version of MOND in this limit.

\subsection{The two-field limit $m_\times \to 0$}
\label{sec:twofield}

In the opposite limit, $m_\times \to 0$, we cannot neglect the double-curl term multiplied by $1/m_\times^2$.
To deal with this term, we use the Helmholtz decomposition.
One particular implication of this decomposition is that, instead of solving a vector field equation $\vec{X} = 0$ one can equivalently solve the system of equations $\vec{\nabla} \cdot \vec{X} = 0$ and $\vec{\nabla} \times \vec{X} = 0$ with the boundary condition that $\vec{X}$ vanishes at infinity \citep{Griffiths2013}.

We apply this to the $\vec{U}$ equation of motion Eq.~\eqref{eq:eom:U}.
This gives the two equations
\begin{align}
\label{eq:Ueomdecomp:div}
 \Delta \Phi = \vec{\nabla} \left[ \vec{U} \left(1 + \tilde{\mu}\left(\frac{|\vec{U}|}{a_0}\right)\right) \right] \,,
\end{align}
and
\begin{multline}
\label{eq:Ueomdecomp:curl}
 - \vec{\nabla} \times \vec{\nabla} \times \vec{\nabla} \times \vec{U} = \\
  2 m_\times^2 \, \vec{\nabla} \times \left[ \vec{U} \left(1 + \tilde{\mu}\left(\frac{|\vec{U}|}{a_0}\right)\right) \right] \,,
\end{multline}
with the reasonable boundary conditions that $\vec{U}$ and $\vec{\nabla} \Phi$ vanish at infinity.
We also decompose the field $\vec{U}$ itself into a divergence-less and a curl-less part,
\begin{align}
 \label{eq:Udecomp}
 \vec{U} = \vec{U}_\times + \vec{\nabla} \alpha \,,
\end{align}
where $\vec{U}_\times$ is a vector field with $\vec{\nabla} \cdot \vec{U}_\times = 0$ and $\alpha$ is a scalar field.
This Helmholtz decomposition is guaranteed to exist when $\vec{U}$ vanishes at least as fast as $1/r$ for $r \to \infty$.
We assume that to be true for now.
Below we will see that this is justified for physically reasonable solutions.

We now consider the implications of the limit $m_\times \to 0$.
The only equation where $m_\times$ appears is the curl-part of the $\vec{U}$ equation Eq.~\eqref{eq:Ueomdecomp:curl}.
Using the Helmholtz decomposition of $\vec{U}$ Eq.~\eqref{eq:Udecomp}, we find
\begin{align}
 \vec{\nabla} \times \vec{\nabla} \times \vec{\nabla} \times \vec{U}_\times = \mathcal{O}(m_\times^2) \,.
\end{align}
Let's assume that $\vec{U}_\times$ (and not just $\vec{U}$ itself) vanishes at infinity.
Then, we find
\begin{align}
 \label{eq:Ucurl:twofield}
 \vec{U}_\times = \mathcal{O}(m_\times^2) \,,
\end{align}
by repeatedly applying the theorem mentioned at the very beginning of this section.
Thus, in the limit $m_\times \to 0$, the vector field $\vec{U}$ is just the gradient of a scalar field,\footnote{
  This means that the vector field $\vec{A}$ has no curl part in the $m_\times \to 0$ limit.
  Since a vanishing curl part was shown to be problematic in Sec.~\ref{sec:novector} one may wonder if we end up with the same problem here.
  Indeed, when one plugs a vector field $\vec{U}$ without a curl term into Eq.~\eqref{eq:eom:U} one again arrives at the problematic equation Eq.~\eqref{eq:PhiA0cond}.
  However, this is not what happens here.
  The reason is that the $\mathcal{O}(m_\times^2)$ curl part of $\vec{U}$ (see Eq.~\eqref{eq:Ucurl:twofield}) cannot be neglected in Eq.~\eqref{eq:eom:U} because of the $1/m_\times^2$ prefactor of the $\vec{\nabla} \times \vec{\nabla} \times \vec{U}$ term there.
  Instead of Eq.~\eqref{eq:AeqA0} we now have $\vec{\nabla} \Phi = \mathcal{O}(m_\times^0) + \vec{\nabla} \alpha \, (1 + \tilde{\mu})$.
  Without the $\mathcal{O}(1)$ term this leads to Eq.~\eqref{eq:PhiA0cond} and the associated inconsistency.
  The $\mathcal{O}(1)$ term prevents this conclusion and, thus, the inconsistency.
  In other words, the $m_\times \to 0$ limit affects both the curl part of the vector field and the $1/m_\times^2$ prefactor in Eq.~\eqref{eq:eom:U}.
  Both are needed to avoid the inconsistency.
  The same does not happen when one takes the (curl part of the) vector field to zero without also taking $m_\times \to 0$, hence the inconsistency we discussed in Sec.~\ref{sec:novector}.
}
\begin{align}
 \vec{U} = \vec{\nabla} \alpha + \mathcal{O}(m_\times^2) \,.
\end{align}

We can plug this result back into the remaining two equations, namely the divergence-part of the $\vec{U}$ equation of motion Eq.~\eqref{eq:Ueomdecomp:div} and the $\Phi$ equation of motion Eq.~\eqref{eq:eom:Phi},
\begin{subequations}
\begin{align}
 \Delta (\Phi - \alpha) &= 4 \pi G_N f_G (\rho_b + \rho_c) \,, \\
 \vec{\nabla} \left( \tilde{\mu}\left(\frac{|\vec{\nabla} \alpha|}{a_0}\right) \vec{\nabla} \alpha \right) &= \Delta (\Phi - \alpha) \,.
\end{align}
\end{subequations}
This can be rewritten as
\begin{subequations}
\label{eq:eomtwo}
\begin{align}
 \Delta (\Phi - \alpha) &= 4 \pi G_N f_G (\rho_b + \rho_c) \,, \\
 \vec{\nabla} \left( \tilde{\mu}\left(\frac{|\vec{\nabla} \alpha|}{a_0}\right) \vec{\nabla} \alpha \right) &= 4 \pi G_N f_G (\rho_b + \rho_c) \,.
\end{align}
\end{subequations}
This is equivalent to the system of equations Eq.~\eqref{eq:eomA0} obtained by setting the vector field $\vec{A}$ to zero, just with $\vec{\nabla} \varphi$ replaced by $\vec{\nabla} \alpha$.
Thus, the procedure of \cite{Skordis2020} is justified in the $m_\times \to 0$ limit.

For the original equations of motion obtained in \cite{Skordis2020}, Eq.~\eqref{eq:eomA0}, the field $\vec{\nabla} \varphi$ does fall off at least as fast as $1/r$ for physical solutions \citep{Skordis2020, Mistele2023}.
Thus, at least in the limit $m_\times \to 0$, the same is true for $\vec{\nabla} \alpha = \vec{U}$ and the decomposition of $\vec{U}$ given in Eq.~\eqref{eq:Udecomp} does indeed exist.

This shows that the $m_\times \to 0$ limit of AeST reproduces MOND-like behavior as long as the effects of the ghost condensate density are negligible.
Thus, in that sense, AeST reproduces a two-field version of MOND in this limit.

\subsection{Equivalence when curl terms vanish}
\label{sec:algebraic}

The only place where $m_\times$ enters in the equations of motion Eq.~\eqref{eq:eom} is as the prefactor of $\vec{\nabla} \times \vec{\nabla} \times \vec{U}$.
In spherical symmetry, this curl term vanishes.
Thus, in spherical symmetry, the value of $m_\times$ has no effect and, consequently, the single-field and two-field limits that we discussed above must be equivalent.
Indeed, this equivalence holds whenever $\vec{\nabla} \times \vec{\nabla} \times \vec{U}$ vanishes identically, though in practice this rarely happens outside spherical symmetry \citep{Brada1995}.
Here, we demonstrate how this equivalence works in practice.

Suppose we are given a solution where $\vec{\nabla} \times \vec{\nabla} \times \vec{U}$ vanishes identically.
This implies that the curl part of $\vec{U}$ vanishes so that $\vec{U}$ can be written as the gradient of a scalar field, $\vec{U} = \vec{\nabla} \alpha$.
In this case, the equations of motion Eq.~\eqref{eq:eom} have the form
\begin{subequations}
\label{eq:Ueomalgebraic}
\begin{align}
 \label{eq:Ueomalgebraic:Phi}
 \Delta (\Phi - \alpha) &= 4 \pi G_N f_G (\rho_b + \rho_c) \,, \\
 \label{eq:Ueomalgebraic:alpha}
 \vec{\nabla} (\Phi - \alpha) &= \tilde{\mu}\left(\frac{|\vec{\nabla} \alpha|}{a_0}\right) \vec{\nabla} \alpha \,.
\end{align}
\end{subequations}
As expected, these are independent of $m_\times$.
One can algebraically solve Eq.~\eqref{eq:Ueomalgebraic:alpha} for $\vec{\nabla} \alpha$ which gives (see the definition of $\mu$ from Eq.~\eqref{eq:mudef}),
\begin{align}
 \label{eq:Ueomalgebraic:alphasolved}
 \vec{\nabla} \alpha = \vec{\nabla} \Phi \,\left(1 - f_G\,\mu\left(\frac{|\vec{\nabla} \Phi|}{a_0}\right)\right) \,.
\end{align}
Importantly, this implies that the combinations $\tilde{\mu}(|\vec{\nabla} \alpha|/a_0) \vec{\nabla} \alpha$ and $\mu(|\vec{\nabla} \Phi|/a_0) \vec{\nabla} \Phi$ are curl-less.
This follows by taking the curl of Eq.~\eqref{eq:Ueomalgebraic:alpha} and Eq.~\eqref{eq:Ueomalgebraic:alphasolved}.

We will now show explicitly that, for such curl-less fields, the full equations of motion Eq.~\eqref{eq:Ueomalgebraic}, the single-field equations Eq.~\eqref{eq:eomsingle}, and the two-field equations Eq.~\eqref{eq:eomtwo} are equivalent.
First, since $\mu(|\vec{\nabla} \Phi|/a_0) \vec{\nabla} \Phi$ is curl-less, the single-field equation Eq.~\eqref{eq:eomsingle} becomes
\begin{align}
 \label{eq:algebraic:single}
 \mu(|x|) \, x = y \,.
\end{align}
Similarly, the two-field equations Eq.~\eqref{eq:eomtwo} become
\begin{subequations}
 \label{eq:algebraic:two}
\begin{align}
 \label{eq:algebraic:two:Newton}
 x - x_\alpha &= f_G \, y \,, \\
 \label{eq:algebraic:two:MOND}
 \tilde{\mu}(|x_\alpha|) \, x_\alpha &= f_G \, y \,.
\end{align}
\end{subequations}
Here, $x = \nabla_N \Phi/a_0$, $x_\alpha = \nabla_N \alpha/a_0$, and $y = |\vec{\nabla} \phi_N|/a_0$ where $\phi_N$ satisfies the equation $\Delta \phi_N = 4 \pi G_N (\rho_b + \rho_c)$ and where a subscript $N$ of a vector denotes its component in the direction of $\vec{\nabla} \phi_N$.
Indeed, whenever a vector $\vec{F}$ is curl-less and satisfies an equation of the form $\vec{\nabla} (\vec{F} - \vec{\nabla} \phi_N) = 0$, then this equation simplifies to $\vec{F} = \vec{\nabla} \phi_N$ or, equivalently, $F_N/a_0 = y$.

The full equations of motion Eq.~\eqref{eq:Ueomalgebraic} (not assuming the single-field or two-field limit) can also be written using this notation.
One finds $x - x_\alpha  = f_G \, y$ and $x - x_\alpha = \tilde{\mu}(|x_\alpha|) x_\alpha$ which is directly seen to be equivalent to the two-field equations Eq.~\eqref{eq:algebraic:two}.
It remains to show that the two-field equations Eq.~\eqref{eq:algebraic:two} are equivalent to the single-field equation Eq.~\eqref{eq:algebraic:single}.

There are two implications to show.
First, consider a solution $x$, $x_\alpha$ of the two-field equations Eq.~\eqref{eq:algebraic:two}.
We show that $x$ also solves the single-field equation Eq.~\eqref{eq:algebraic:single}.
An important ingredient is the implicit definition of $\mu$ in terms of $\tilde{\mu}$ from Eq.~\eqref{eq:mudef}.
In terms of the notation used here, this implicit definition is based on the equation $x = x_\alpha (1+\tilde{\mu}(|x_\alpha|))$ which follows from Eq.~\eqref{eq:algebraic:two:Newton} and Eq.~\eqref{eq:algebraic:two:MOND}.
Indeed, solving this equation for $x_\alpha$ gives $x_\alpha = x (1 - f_G \mu(|x|))$ by our definition of $\mu$.
Adding up these two relations gives $\tilde{\mu}(|x_\alpha|) x_\alpha = f_G \mu(|x|) x$.
Using Eq.~\eqref{eq:algebraic:two:MOND} then gives $\mu(|x|) x = y$.
This is to say that $x$ solves the single-field equation Eq.~\eqref{eq:algebraic:single}, which is what was to be shown.

Next, consider a solution $x$ of the single-field equation Eq.~\eqref{eq:algebraic:single}.
We show that this $x$ is a solution of the two-field equations Eq.~\eqref{eq:algebraic:two}.
We define a candidate $x_\alpha$ as $x_\alpha \equiv x - f_G y$.
By construction, this solves Eq.~\eqref{eq:algebraic:two:Newton}.
It remains to show that it also solves Eq.~\eqref{eq:algebraic:two:MOND}.
To this end, we again use the implicit definition of $\mu$ in terms of $\tilde{\mu}$ which again gives $\tilde{\mu}(|x_\alpha|) x_\alpha = f_G \mu(|x|) x$.
Using Eq.~\eqref{eq:algebraic:single} then implies $\tilde{\mu}(|x_\alpha|) x_\alpha = f_G y$ which is Eq.~\eqref{eq:algebraic:two:MOND}.
Thus, the algebraic single-field equation Eq.~\eqref{eq:algebraic:single} and the algebraic two-field equations Eq.~\eqref{eq:algebraic:two} are equivalent.
That is, the single-field and two-field limits are equivalent when the curl part of $\vec{U}$ vanishes identically.

For later reference, we note that these algebraic equations are often a reasonable approximation in the context of MOND even when the curl terms discussed here do not vanish identically \citep{Famaey2012}.
Writing the solution of the equation $\mu(|x|) x = y$ in the form $x = \nu(y) y$ one has in this approximation
\begin{align}
 \vec{a}_{\mathrm{tot}} = \vec{a}_N \, \nu\left(\frac{|\vec{a}_N|}{a_0}\right) \,,
\end{align}
where $\vec{a}_N$ is the Newtonian acceleration due to the mass corresponding to the density $\rho \equiv \rho_b + \rho_c$ which includes both baryonic matter and the ghost condensate, i.e. $\vec{a}_N = - \vec{\nabla} \phi_N$.
Thus, in the case where the ghost condensate is negligible, i.e. when $\vec{a}_N \approx \vec{a}_b$, this gives the standard MOND relation
\begin{align}
 \label{eq:algebraic}
  \vec{a}_{\mathrm{tot}} = \vec{a}_b \, \nu(|\vec{a}_b|/a_0) \,.
\end{align}

\section{Scale-dependence of the two limits}
\label{sec:scale}

Above, we have seen that AeST reduces to a single-field version of MOND for $m_\times \to \infty$ and to a two-field version for $m_\times \to 0$.
Both limits reproduce MOND-like behavior to the same extent as the equations Eq.~\eqref{eq:eomA0} originally derived in Ref.~\cite{Skordis2020} assuming a vanishing vector field.
That is, both limits reproduce MOND as long as the ghost condensate is negligible.
Thus, there are two scales to consider:
The model parameter $m_\times$ controls whether the single-field limit or the two-field limit applies, and the model parameter $m$ (that multiplies the ghost condensate density $\rho_c$) controls whether these limits reproduce MOND \citep{Skordis2020,Mistele2023,Verwayen2023}.

In practice, whether the single-field or the two-field limit is applicable depends not only on the value of the model parameter $m_\times$.
It also depends on the system under consideration.
In particular, consider a system with typical length scale $l$.
Then, derivatives can very roughly be estimated to be of order $1/l$.
Thus, in practice, the relevant quantity is $m_\times l$ (see Eq.~\eqref{eq:eom:U}),
\begin{align}
 m_\times l \gg 1&: \quad \mathrm{single-field\;limit} \,, \\
 m_\times l \ll 1&: \quad \mathrm{two-field\;limit} \,.
\end{align}
That is, even after fixing the AeST model parameters, some systems will behave as in the single-field limit while others will behave as in the two-field limit.
It should be kept in mind, however, that this distinction is irrelevant in spherically symmetric situations where these two limits give equivalent equations.

The quantity $m_\times = \sqrt{(2-K_B)/K_B} Q_0$ is, to the best of our knowledge, not yet constrained phenomenologically.
For the explicit examples of model parameters given in \cite{Skordis2020}, we have
\begin{align}
 \label{eq:mxvalues}
 10^{-4}\,\mathrm{Mpc}^{-1} \lesssim m_\times \lesssim 1\,\mathrm{Mpc}^{-1} \,.
\end{align}
Thus, in the following, we will assume this range of parameter values, though other choices are certainly possible.
For this range of parameter values, the transition from the two-field limit (small spatial scales) to the single-field limit (large spatial scales) happens for $l \sim \mathrm{Mpc}$ or larger.

To be concrete, this means that the two-field limit is relevant for wide binaries ($l \sim \mathrm{kAU}$), and at least the inner parts of galaxies ($l \sim \mathrm{kpc}$).
For galaxy clusters ($l \sim \mathrm{Mpc}$), the single-field limit may be relevant depending on the details of the cluster under consideration and the precise value of $m_\times$.

As discussed above, the single-field and two-field limits both recover MOND-like behavior when the effects of the ghost condensate are negligible.
Typically, the ghost condensate becomes important beyond a critical radius $r_c$ whose scale is set by $m$ \citep{Skordis2020,Mistele2023,Verwayen2023},
\begin{align}
 l \gtrsim \left( \frac{r_M}{m^2/f_G} \right)^{1/3} \equiv r_c \,.
\end{align}
Here, $r_M \equiv \sqrt{G M_b / a_0}$ with the baryonic mass of the system $M_b$.
A typical order of magnitude of $m$ is $1\,\mathrm{Mpc}^{-1}$ \citep{Skordis2020,Skordis2021}.
Thus, $m$ and $m_\times$ can be comparable and -- depending on the values of $m$, $m_\times$, $l$ and $M_b$ -- there are four possible regimes, corresponding to the four possible combinations of single-field/two-field limit and significant/negligible $\rho_c$.

For the explicit examples we consider below in Sec.~\ref{sec:mw} and Sec.~\ref{sec:wb} -- galactic rotation curves and wide binaries -- we expect that the two-field limit applies ($m_\times l \ll 1$) and that the ghost condensate is unimportant ($l \ll r_c$).
Indeed, with $m \sim 1\,\mathrm{Mpc}^{-1}$ and with $m_\times$ in the range given by Eq.~\eqref{eq:mxvalues}, the single-field limit ($l \gg m_\times^{-1}$) may never be relevant when AeST reproduces MOND ($l \ll r_c$).
Below, we nevertheless also show the results for the single-field limit because this illustrates the difference to other models of MOND for which the single-field equation Eq.~\eqref{eq:eomsingle} is relevant (e.g. Ref.~\cite{Bekenstein1984}) and also because (in contrast to $m$ \citep{Skordis2021,Mistele2023,Verwayen2023}) the value of $m_\times$ is so far not constrained by observations so that choices other than Eq.~\eqref{eq:mxvalues} are, at least in principle, possible.

\section{Galactic rotation curves are almost the same in both limits}
\label{sec:mw}

Above, we have seen that the quasi-static limit of AeST is more complicated than what was proposed in \cite{Skordis2020} based on the assumption of a vanishing vector field $\vec{A}$.
The procedure of \cite{Skordis2020} is justified for systems much smaller than the model parameter $1/m_\times$, but not in general.
In practice, however, the value of $m_\times$ is often unimportant.
That is, in practice, one can often follow the procedure of \cite{Skordis2020} even for systems larger than $1/m_\times$.

\begin{figure}
 \centering
 \includegraphics[width=\columnwidth]{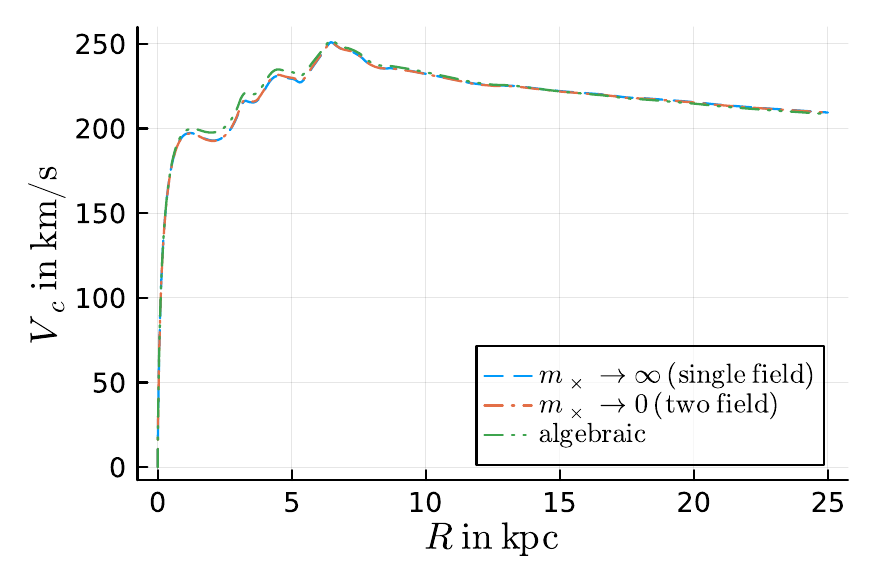}
 \caption{
   The rotation curve of the Milky Way for the single-field limit $m_\times \to 0$ (dashed blue line), the two-field limit $m_\times \to 0$ (dash-dotted red line), and as calculated from the algebraic MOND relation Eq.~\eqref{eq:algebraic}.
   This is with the simple interpolation function $\mu(x) = x/(1+x)$, $f_G = 0.99$, and without the ghost condensate mass, i.e. for $m^2/f_G = 0$.
   We assume the Milky Way mass model from \cite{McGaugh2019b}.
 }
 \label{fig:mw-compare-vc}
\end{figure}

As discussed in Sec.~\ref{sec:algebraic}, one particular example is spherical symmetry.
Many astrophysically relevant systems are not spherically symmetric, of course.
But even in axisymmetric systems does the value of $m_\times$ not matter in practice.
This can be seen from Fig.~\ref{fig:mw-compare-vc}, which shows the Milky Way rotation curve as calculated for the two limits $m_\times \to 0$ and $m_\times \to \infty$ for the Milky Way mass model from \cite{McGaugh2019b}.
Here, we assume $m^2/f_G = 0$ for simplicity.
That is, we assume that the ghost condensate density vanishes.
This is a good approximation at the radii considered here, at least for the usual choice $\sqrt{f_G}/m \gtrsim \mathrm{Mpc}$ \citep{Skordis2020, Mistele2023,Verwayen2023}.
Details of the numerical procedure are described in Appendix~\ref{sec:appendix:numerics}.

For reference, Fig.~\ref{fig:mw-compare-vc} also shows a rotation curve obtained from the algebraic relation Eq.~\eqref{eq:algebraic}.
This algebraic relation is often considered a reasonable approximation in MOND, even outside spherical symmetry.
We see that the difference between the $m_\times \to 0$ and $m_\times \to \infty$ limits is much smaller than that between the algebraic case and either value of $m_\times$.
Indeed, the effect of $m_\times$ is much smaller than the systematic uncertainties in modelling the Milky Way \citep{McGaugh2019b}.
Thus, in practice, the value of $m_\times$ is unlikely to play a role for galactic rotation curves.

\begin{figure}
 \centering
 \includegraphics[width=\columnwidth]{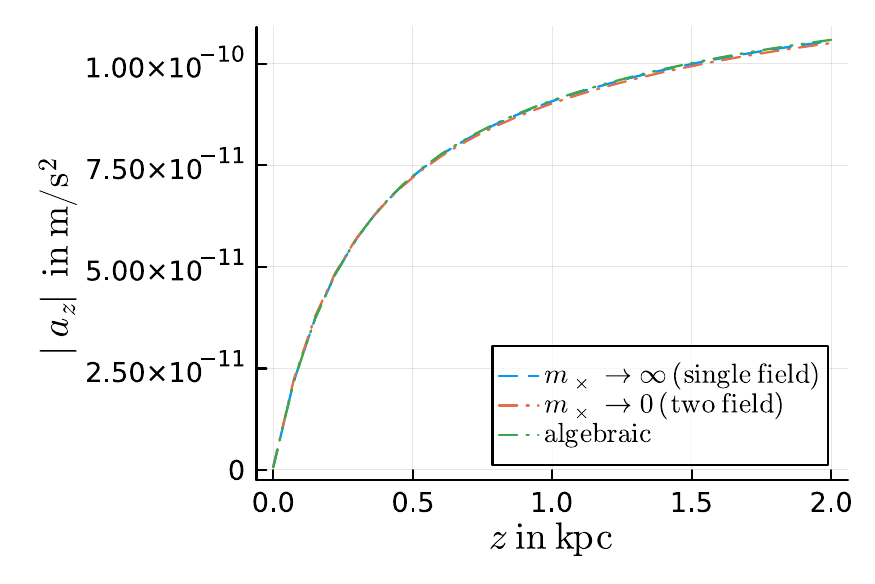}
 \caption{
   The vertical acceleration for the same Milky Way model as shown in Fig.~\ref{fig:mw-compare-vc} at the solar radius $R = 8.1\,\mathrm{kpc}$ for the single-field limit $m_\times \to \infty$ (dashed blue line), the two-field limit $m_\times \to 0$ (dash-dotted red line), and as calculated from the algebraic MOND relation from Eq.~\eqref{eq:algebraic}.
 }
 \label{fig:mw-compare-az}
\end{figure}

The same holds for observables based on the vertical acceleration above the Milky Way disk (see, e.g., \cite{Lisanti2019, Zhu2023}).
This is illustrated in Fig.~\ref{fig:mw-compare-az} which shows almost no difference between the limits $m_\times \to 0$ and $m_\times \to \infty$.

\section{Percent-level effect for wide binaries}
\label{sec:wb}

The field equations of MOND are inherently non-linear.
As a result, the environment of a system can affect its internal dynamics.
This is the so-called External Field Effect (EFE) and it violates the Strong Equivalence Principle.
A version of this EFE also exists in AeST because its field equations share the same type of non-linearity.
For us, the important thing is that the single-field limit $m_\times \to \infty$ and the two-field limit $m_\times \to 0$ of AeST do not produce the same EFE \citep{Milgrom2009b}.

One particular type of system where the EFE is important are wide binaries \citep[see, for example, ][]{Hernandez2012,Banik2015,Banik2018,Pittordis2019,Hernandez2022}.
Intriguingly, recent observational studies of wide binaries are getting close to percent-level accuracy in acceleration \citep{Pittordis2023,Hernandez2023,Chae2023,Chae2023b,Banik2023}.
Thus, even small differences in what AeST predicts compared to other models can be important.

As discussed in Sec.~\ref{sec:scale}, wide binaries satisfy $m_\times l \ll 1$ and $l \ll r_c$.
Thus, the relevant limit of AeST is the two-field limit $m_\times \to 0$ with a negligible ghost condensate density $\rho_c$.
Previous works have not considered this two-field version of MOND, focusing on other models such as that obtained in the $m_\times \to \infty$ limit of AeST \citep[i.e. AQUAL,][]{Banik2015, Banik2018}.
Thus, any difference between these two limits indicates that the AeST prediction for wide binaries deviates from what previous studies considered.
In this regard, a more general point is that different models that reproduce the same basic tenets of MOND, such as being able to explain the RAR, may differ in their secondary predictions such as the size of the EFE \citep{Milgrom2014, Milgrom2023}.

As we will now show, there can indeed be percent-level differences between the single-field and and two-field limits of AeST.
However, there's also reason to be careful with quantitative statements.
As we will argue, in the two-field limit, wide binaries are much more sensitive to the details of the local Galactic field than in the single-field limit.
Any quantitative statement about the difference between these two limits must be understood to be relative to a particular model of the Milky Way and it is plausible that different Milky Way models erase or significantly enhance the percent-level differences mentioned above.

To see this, we again assume that the ghost condensate's mass is not important or, equivalently, that $m^2/f_G = 0$.
This is a good approximation on the small scales, $l \sim \mathrm{kAU}$, that are relevant here.
We also assume that the EFE dominates everywhere in the wide binary system under consideration.
This approximation is reasonable when the distance between the two stars in the binary system is relatively large \citep{Banik2015, Banik2018}.
Importantly, these approximations allow us to make analytical estimates.
More involved numerical calculations are left for future work.

As already mentioned, for the internal dynamics of wide binaries, the two-field limit $m_\times \to 0$ is the relevant one.
For simplicity, we assume that the external field produced by the Milky Way can be calculated using the same limit.
We expect this to be a good approximation, since the $m_\times \to 0$ limit is valid up to scales of order $1/m_\times$ which in our case means up to $1\,\mathrm{Mpc} - 10^4\,\mathrm{Mpc}$.
Thus, we generally expect only sub-percent differences between the actual external field and that calculated using the $m_\times \to 0$ limit.\footnote{
  In Appendix~\ref{sec:appendix:generalefe}, we consider what happens when the external and internal fields cannot be calculated using the same assumption about whether $m_\times l$ is large or small.
  We show that, in this case, the external fields $x_e$ and $x_{\varphi,e}$ (see below) have to be calculated differently but Eq.~\eqref{eq:aoveraNsingle} and Eq.~\eqref{eq:aoveraNtwo} remain valid.
  Thus, any deviations from our results due to this can be captured in a factor $f$ as introduced below in Eq.~\eqref{eq:f}.
}

For reference, we first consider the single-field limit $m_\times \to \infty$, assuming that the external field is also calculated in this limit.
This is equivalent to AQUAL and is explicitly discussed in \cite{Banik2015}.
They find for the radial acceleration in the binary system, $a$, relative to that in the Newtonian case, $a_N$,
\begin{align}
 \label{eq:aoveraNsingle}
 \left. \frac{a}{a_N} \right|_{m_\times \to \infty} = \frac{1}{\mu(x_e) \sqrt{1 + \sin^2 \theta \, L(x_e)}} \,.
\end{align}
Here, $\mu(x)$ is  the interpolation function of single-field MOND (see Sec.~\ref{sec:singlefield}), $L(x)$ is its logarithmic derivative,
\begin{align}
 L(x) \equiv \frac{x \mu'(x)}{\mu(x)} \,,
\end{align}
$x_e$ is the gradient of the external gravitational potential $\Phi_e$ relative to $a_0$,
\begin{align}
 x_e = \frac{|\vec{\nabla} \Phi_e|}{a_0} \,,
\end{align}
and $\theta$ is the angle between the radial direction (connecting the two stars in the binary system) and the external field.

It is straightforward to adopt the procedure of \cite{Banik2015} to the two-field limit $m_\times \to 0$.
We find
\begin{align}
 \label{eq:aoveraNtwo}
 \left. \frac{a}{a_N} \right|_{m_\times \to 0} = f_G \left(1 + \frac{1}{\tilde{\mu}(x_{\varphi,e}) \sqrt{1 + \sin^2 \theta \, \tilde{L}(x_{\varphi,e})}} \right) \,,
\end{align}
where $\tilde{\mu}(x)$ is the two-field version of the interpolation function (see Sec.~\ref{sec:twofield}), $\tilde{L}(x)$ is its logarithmic derivative, and $x_{\varphi,e}$ is the gradient of the external gravitational potential $\varphi_e$ relative to $a_0$,
\begin{align}
 x_{\varphi,e} = \frac{|\vec{\nabla} \varphi_e|}{a_0} \,.
\end{align}
This expression for $a/a_N$ has a very similar structure as the corresponding single-field expression Eq.~\eqref{eq:aoveraNsingle}, except there are now two terms, corresponding to the two fields $\Phi$ and $\varphi$, the single-field interpolation function $\mu$ is replaced with its two-field counterpart $\tilde{\mu}$, and there is an overall factor of $f_G$.

Importantly, in the two-field limit, it is only the external field of the field $\varphi$ that plays a role, the external field of $\Phi$ is irrelevant.
This is because the $\Phi$ equation of motion is linear so that there is no EFE.

This is related to an important complication in the two-field limit, regarding how to choose the value of $x_{\varphi,e}$.
In the single-field limit, a simple way to choose the external gravitational field is to reproduce the observed circular rotation curve of the Milky Way at the solar radius,
\begin{align}
 x_e = \frac{|\vec{\nabla} \Phi_e|}{a_0} \equiv \frac{1}{a_0} \frac{V_c^2}{R_\odot} \,.
\end{align}
In the two-field limit, reproducing the value of $V_c^2/R_\odot$ means
\begin{align}
 x_{\Phi,e} + x_{\varphi,e} = |\vec{\nabla} \Phi_e + \vec{\nabla} \varphi_e| \equiv \frac{1}{a_0} \frac{V_c^2}{R_\odot} \,,
\end{align}
where $x_{\Phi, e} \equiv |\vec{\nabla} \Phi_e|/a_0$ and, for simplicity, we assumed that $\vec{\nabla} \Phi_e$ and $\vec{\nabla} \varphi_e$ point in the same direction.
The important point is that, in the single-field case, the observed value of $V_c^2/R_\odot$ directly gives $x_e$ and that is all that is needed to calculate $a/a_N$ from Eq.~\eqref{eq:aoveraNsingle}.
This is different in the two-field case.
The observed value of $V_c^2/R_\odot$ only gives the sum $x_{\Phi,e} + x_{\varphi,e}$ while calculating $a/a_N$ from Eq.~\eqref{eq:aoveraNtwo} requires the $x_{\varphi,e}$ part of that sum individually.

This is the complication for the two-field case mentioned above.
It requires additional assumptions about the external field to know which part of the total acceleration $V_c^2/R_\odot$ comes from $\Phi$ and which part comes from $\varphi$.

Here, we first consider a spherically symmetric external field.
Of course, the Milky Way is not spherically symmetric,  but this case serves as an illustrative starting point.
We discuss more general cases below.
Assuming spherical symmetry, the total acceleration $x_{\mathrm{tot},e} \equiv |\vec{\nabla} \Phi_e + \vec{\nabla} \varphi_e|/a_0$ in the two-field case satisfies the standard algebraic MOND relation with the interpolation function $\mu$ from the single-field limit (see also Sec.~\ref{sec:algebraic}),
\begin{align}
\label{eq:xtotalgebraic}
\mu(x_{\mathrm{tot},e}) \, x_{\mathrm{tot},e} = y_e \,,
\end{align}
where $y_e$ is the gradient of the Newtonian external potential relative to $a_0$.
This follows from Eq.~\eqref{eq:eomtwo} in spherical symmetry and the definition of $\mu$ in terms of $\tilde{\mu}$ Eq.~\eqref{eq:mudef}.
In addition, we have $x_{\Phi,e} = f_G \, y_e$ directly from Eq.~\eqref{eq:eomtwo}.
This gives
\begin{align}
 \label{eq:xvarphie:spherical}
 x_{\varphi,e} = x_\odot ( 1 - f_G \mu(x_\odot)) \, \quad \mathrm{with} \quad x_\odot \equiv \frac{1}{a_0} \frac{V_c^2}{R_\odot} \,.
\end{align}

\begin{figure}
 \centering
 \includegraphics[width=\columnwidth]{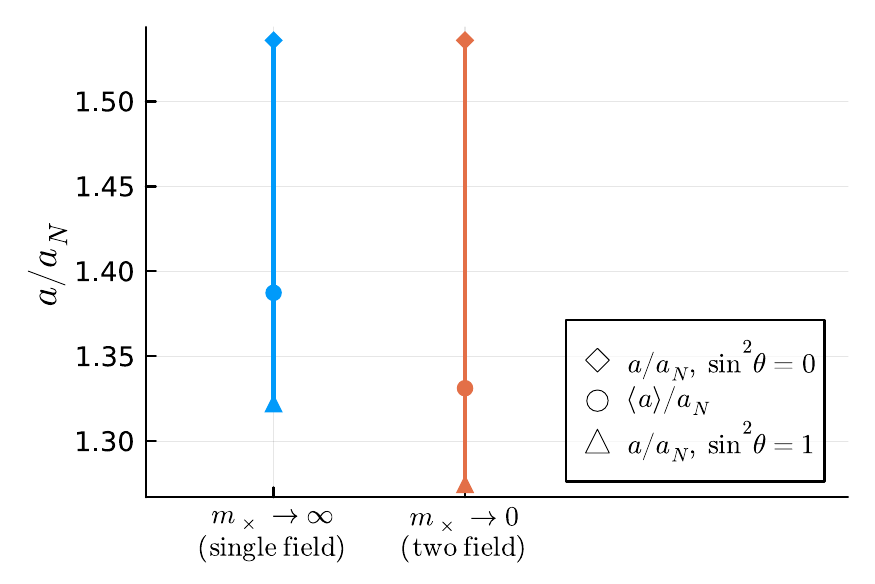}
 \caption{
   The radial acceleration in a wide binary system relative to that in the Newtonian case for the single-field limit $m_\times \to \infty$ (blue) and the two-field limit $m_\times \to 0$ (red) of AeST.
   For real-world wide binaries with separations $l \sim \mathrm{kAU}$, the two-field limit $m_\times \to 0$ is the relevant one.
   Markers with a diamond shape indicate an external field parallel to the radial direction, $\sin^2 \theta = 0$, markers with a triangle shape indicate a perpendicular external field, $\sin^2 \theta = 1$, and markers with a circle shape indicate the angular average from Eq.~\eqref{eq:aoveraNaverage}.
   We assume that the external field dominates everywhere in the binary system and use the simple interpolation function $\mu(x) = x/(1+x)$.
   We further assume $f_G = 0.99$ and no ghost condensate mass, i.e. $m^2/f_G = 0$.
   The external field is calculated from the acceleration $V_c^2/R$ at the solar radius $R = 8.1\,\mathrm{kpc}$ for the Milky Way model shown in Fig.~\ref{fig:mw-compare-vc}.
   For the single-field limit, the value of $V_c^2/R$ alone suffices to calculate $a/a_N$.
   For the two-field limit, we additionally assume the algebraic MOND relation Eq.~\eqref{eq:algebraic}.
 }
 \label{fig:wb-compare-algebraic}
\end{figure}

Thus, assuming spherical symmetry, we can calculate $a/a_N$ from just $V_c^2/R_\odot$ even in the two-field limit.
Explicitly, we have
\begin{align}
 \left. \frac{a}{a_N} \right|^{\mathrm{spherical}}_{m_\times \to 0} = f_G \left(1 + \frac{1}{\tilde{\mu}(x_{\varphi,e}) \sqrt{1 + \sin^2 \theta \, \tilde{L}(x_{\varphi,e})}} \right)\,,
\end{align}
with $x_{\varphi,e} = x_\odot ( 1 - f_G \mu(x_\odot))$ and $x_\odot$ as defined in Eq.~\eqref{eq:xvarphie:spherical}.
This should be compared to the single-field result Eq.~\eqref{eq:aoveraNsingle} with $x_e = x_\odot$.
We quantitatively compare these two cases in Fig.~\ref{fig:wb-compare-algebraic}.

From Fig.~\ref{fig:wb-compare-algebraic}, we see that the two-field and the single-field limits give the same acceleration when the external field points in the radial direction, $\sin^2 \theta = 0$.
In contrast, when the external field is perpendicular, $\sin^2 \theta = 1$, the acceleration in the two-field limit is a few percent smaller than in the single-field limit.
Following \cite{Banik2018}, one may also consider an acceleration averaged over $\theta$,
\begin{align}
 \label{eq:aoveraNaverage}
 \langle a \rangle \equiv \frac12 \int_0^{\pi} d \theta \sin \theta \, a \,.
\end{align}
As we can see from Fig.~\ref{fig:wb-compare-algebraic}, this averaged acceleration is again smaller for the two-field limit than for the single-field limit.
The magnitude of the effect after averaging is similar to that in the perpendicular case without averaging.

These results are often a good approximation even beyond spherical symmetry, namely whenever the algebraic relation $\mu(x_{\mathrm{tot},e}) \, x_{\mathrm{tot},e} = y_e$ from Eq.~\eqref{eq:xtotalgebraic} is a good approximation.
This is indeed often the case (see, e.g., \cite{Bekenstein1984, Hossenfelder2020}).

However, one also must be careful here.
Since the difference between the single-field and the two-field limits is just a few percent, one should use the algebraic relation Eq.~\eqref{eq:xtotalgebraic} only if it is valid well below the percent level.
Otherwise, any real effect may be washed out by errors in the external field.

\begin{figure}
 \centering
 \includegraphics[width=\columnwidth]{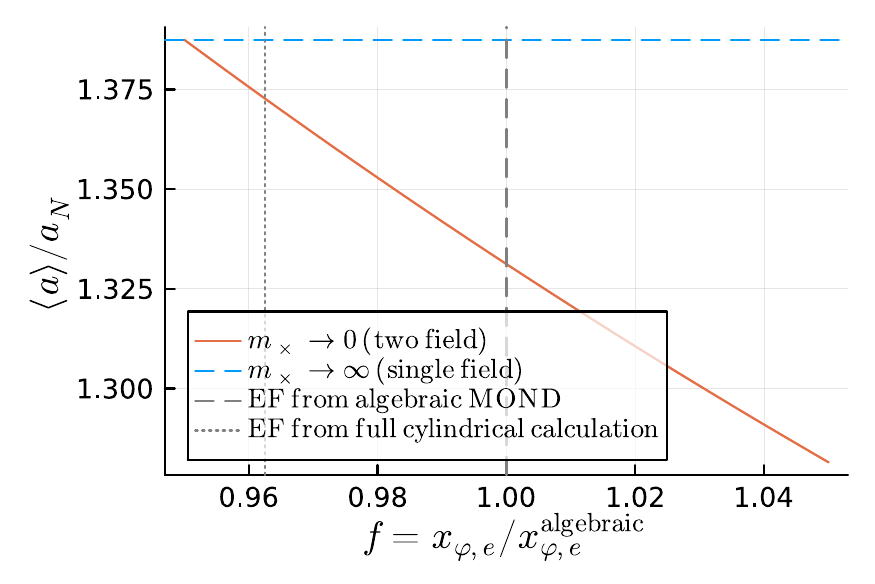}
 \caption{
   The angle-averaged acceleration Eq.~\eqref{eq:aoveraNaverage} in a wide  binary system relative to the Newtonian case as a function of how much the external field $x_{\varphi,e}$ in the two-field limit $m_\times \to 0$ deviates from what the algebraic relation Eq.~\eqref{eq:algebraic} implies (solid red line).
   For comparison we also show the single-field limit $m_\times \to \infty$ (dashed blue line) which does not depend on $f$.
   Apart from this factor $f$, the calculation is the same as in Fig.~\ref{fig:wb-compare-algebraic}.
   The vertical dashed line denotes the prediction for an external field that follows the algebraic MOND relation, $f=1$, while the dotted vertical line shows what we find for the axisymmetric Milky Way model from \cite{McGaugh2019b}, see also Fig.~\ref{fig:mw-compare-vc}.
 }
 \label{fig:wb-compare-uMfactor}
\end{figure}

We show the effect of deviations from the algebraic relation Eq.~\eqref{eq:xtotalgebraic} in Fig.~\ref{fig:wb-compare-uMfactor}.
In particular, we show how the averaged acceleration ratio $\langle a \rangle / a_N$ in the two-field limit changes when the gradient of the external field $x_{\varphi,e}$ deviates from what the algebraic relation Eq.~\eqref{eq:xtotalgebraic} says it should be.
We parametrize the deviation of $x_{\varphi,e}$ from its value derived from this algebraic relation by a factor $f$,
\begin{align}
 \label{eq:f}
 f \equiv \frac{x_{\varphi,e}}{x_{\varphi,e}^{\mathrm{algebraic}}} \equiv \frac{x_{\varphi,e}}{x_\odot(1 - f_G \mu(x_\odot))} \,.
\end{align}
From Fig.~\ref{fig:wb-compare-uMfactor}, we see that a deviation of $x_{\varphi,e}$ from its algebraic value by a few percent also changes the acceleration $\langle a \rangle/a_N$ by a few percent.
In particular, smaller values of $x_{\varphi,e}$ bring the single-field and the two-field limit closer together, while larger values push them apart.
A moderately smaller value, $f \approx 0.95$, erases the difference between the two limits.

Thus, compared to the single-field limit, one needs to know the details of the Milky Way's gravitational field much more precisely when making predictions for wide binaries in the two-field limit.
As one particular example, we here consider the axisymmetric Milky Way model already discussed in Sec.~\ref{sec:mw}.
In that model, comparing the numerically computed value of $x_{\varphi,e}$ at the solar radius to that inferred from the algebraic relation Eq.~\eqref{eq:xtotalgebraic} gives
\begin{align}
 f \approx 0.963 \,.
\end{align}
This value is also highlighted in Fig.~\ref{fig:wb-compare-uMfactor}.
We see that this leaves a difference of only about $1.5$ percentage points between the single-field and the two-field limit.

One might conclude that the value of $m_\times$ makes very little difference for wide binaries.
That is, one might expect that the AeST prediction (corresponding to the two-field limit $m_\times \to 0$) is very close to what previous studies have found for AQUAL (i.e. for the single-field limit $m_\times \to \infty$).
However, as we have seen, any quantitative statement depends sensitively on the details of the local gravitational field of the Milky Way.
Indeed, the Milky Way disk is found not to be in equilibrium and non-axisymmetries may not be negligible \citep{Antoja2018,Faure2014,McGaugh2019b}.
Plausibly, such effects can either erase or significantly enhance the percent-level differences between AeST and other models of MOND.
Thus, quantitative predictions require more involved numerical studies that take such effects into account.
They also require carefully evaluating how accurate such predictions can be given the observational uncertainties.
This is left for future work.

\section{Phenomenology of curl terms}
\label{sec:curlpheno}

Above, we have seen that the single-field limit $m_\times \to \infty$ and the two-field limit $m_\times \to 0$ produce practically indistinguishable rotation curves in galaxies but show larger differences in wide binaries.
Here, we show that the larger effect of $m_\times$ in wide binaries is due to curl terms that are closely related to (but not identical with) the curl part of the vector field $\vec{U}$.

We first note that the curl part of $\vec{U}$ is not the only relevant curl in the quasi-static limit of AeST.
Indeed, the equations of motion in the single-field limit, Eq.~\eqref{eq:eomsingle}, and in the two-field limit, Eq.~\eqref{eq:eomtwo}, are both of the form $\vec{\nabla} \vec{F} = 4 \pi G \rho$.
This is equivalent to $\vec{\nabla} (\vec{F} - \vec{\nabla} \phi_N)$ where $\phi_N$ satisfies the Newtonian Poisson equation $\Delta \phi_N = 4\pi G \rho$.
In general, this implies $\vec{F} = \vec{\nabla} \phi_N + \vec{\nabla} \times \vec{h}$ for some vector field $\vec{h}$, i.e. $\vec{F}$ matches $\vec{\nabla} \phi_N$ up to a curl term $\vec{\nabla} \times \vec{h}$.
As we will see, curl terms like $\vec{\nabla} \times \vec{h}$ are directly responsible for the fact that $m_\times$ has a larger effect in wide binaries than in galaxies.

Specifically, in the single-field limit $m_\times \to \infty$ we have
\begin{align}
 \label{eq:curlterm:single}
 \mu\left(\frac{|\vec{\nabla} \Phi|}{a_0}\right) \vec{\nabla} \Phi = \vec{\nabla} \phi_N + \vec{\nabla} \times \vec{h}_\infty \,,
\end{align}
for some vector field $\vec{h}_\infty$.
Similarly, in the two-field limit $m_\times \to 0$, we have
\begin{subequations}
\label{eq:curlterm:two}
\begin{align}
\label{eq:curlterm:two:Newton}
\vec{\nabla} (\Phi - \alpha) &= f_G \vec{\nabla} \phi_N \,, \\
\tilde{\mu}\left(\frac{|\vec{\nabla} \alpha|}{a_0}\right) \vec{\nabla} \alpha &= f_G \vec{\nabla} \phi_N + \vec{\nabla} \times \vec{h}_0 \,,
\end{align}
\end{subequations}
for some vector field $\vec{h}_0$.
There is no curl term for Eq.~\eqref{eq:curlterm:two:Newton} since any potential curl term $\vec{\nabla} \times \vec{h}$ that one might add to Eq.~\eqref{eq:curlterm:two:Newton} can be shown to vanish by taking the curl on both sides.

As discussed in Sec.~\ref{sec:algebraic}, when these curl terms vanish, the single-field limit and the two-field limit are equivalent, i.e. both limits produce the same total acceleration felt by matter $\vec{a}_{\mathrm{tot}} = - \vec{\nabla} \Phi$.
In this case, one has
\begin{align}
 \label{eq:algebraic-fields}
 \left. \vec{\nabla} \Phi\right|_{\mathrm{no\,curl}} = \nu\left(\frac{|\vec{\nabla} \phi_N|}{a_0}\right) \vec{\nabla} \phi_N \,,
\end{align}
irrespective of the value of $m_\times$.
Thus, any difference between the $m_\times \to 0$ and $m_\times \to \infty$ limits must be due to non-zero curl terms such as $\vec{\nabla} \times \vec{h}_\infty$ and $\vec{\nabla} \times \vec{h}_0$.

Indeed, when these curl terms do not vanish, Eq.~\eqref{eq:algebraic-fields} receives corrections that we parametrize in terms of a vector field $\vec{C}$, defined by
\begin{align}
 \label{eq:curlterm:C}
 \vec{\nabla} \Phi \equiv \nu\left(\frac{|\vec{\nabla} \phi_N|}{a_0}\right) \vec{\nabla} \phi_N - \vec{C}\,.
\end{align}
Given a curl term $\vec{\nabla} \times \vec{h}_\infty$ (in the single-field limit) or $\vec{\nabla} \times \vec{h}_0$ (in the two-field limit) it is straightforward to obtain the corresponding $\vec{C}$ by first algebraically solving Eq.~\eqref{eq:curlterm:single} or Eq.~\eqref{eq:curlterm:two} for $\vec{\nabla} \Phi$ and then subtracting $\vec{\nabla} \Phi$ from $\nu(|\vec{\nabla} \phi_N|/a_0) \vec{\nabla} \phi_N$.
In the following we will mostly consider $\vec{C}$ instead of $\vec{\nabla} \times \vec{h}_\infty$ and $\vec{\nabla} \times \vec{h}_0$ because $\vec{C}$ is more directly related to observations.
Indeed, $\vec{C}$ is the additional acceleration induced by the curl terms.
We refer to the rest of the acceleration (not induced by curl terms) as the algebraic part $\vec{a}_{\mathrm{algebraic}}$ of the total acceleration, i.e. $\vec{a}_{\mathrm{tot}} = \vec{a}_{\mathrm{algebraic}} + \vec{C}$.

\begin{figure}
 \centering
 \includegraphics[width=\columnwidth]{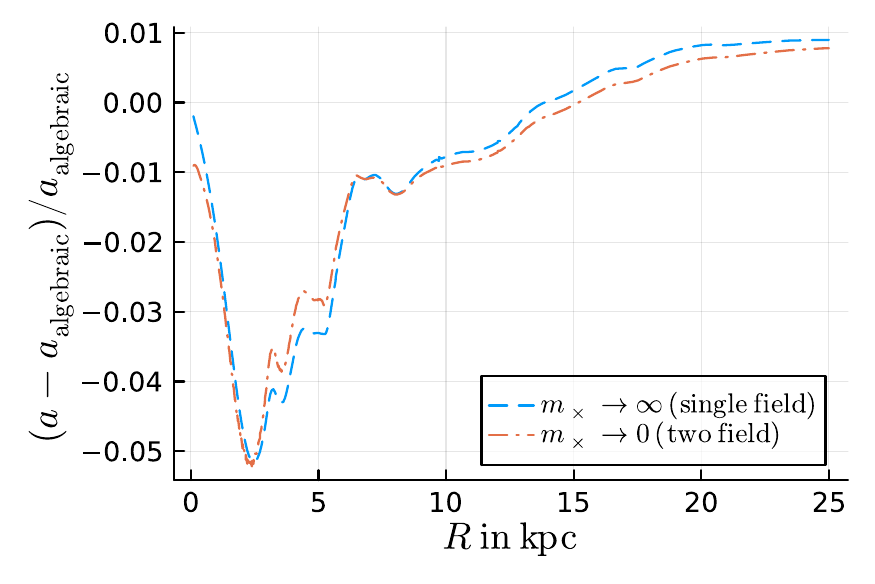}
 \caption{
   The acceleration induced by curl terms in the single-field limit $m_\times \to \infty$ (dashed blue line, see Eq.~\eqref{eq:curlterm:single}) and the two-field limit $m_\times \to 0$ (dash-dotted red line, see Eq.~\eqref{eq:curlterm:two}) for the Milky Way model from Sec.~\ref{sec:mw} at $z=0$.
   These curl-induced accelerations are why, in general, the single-field and two-field limits are not equivalent.
   The part of the total acceleration that is not induced by curl terms, $\vec{a}_{\mathrm{algebraic}}$, is independent of $m_\times$ (see Sec.~\ref{sec:algebraic}).
   For the Milky Way model considered here, the curl-induced accelerations in the single-field and two-field limits are almost identical which is why the corresponding rotation curves are almost identical in both limits (see Fig.~\ref{fig:mw-compare-vc}).
 }
 \label{fig:mw-curl-terms}
\end{figure}

Fig.~\ref{fig:mw-curl-terms} shows the acceleration induced by the curl terms, $\vec{C}$, relative to the algebraic acceleration, $\vec{a}_{\mathrm{algebraic}}$, for the Milky Way model considered in Sec.~\ref{sec:mw} at $z=0$.
We see that the curl terms affect the acceleration on the percent level.
This is what produces the percent-level difference between the rotation curve calculated from the algebraic MOND relation Eq.~\eqref{eq:algebraic} and  the full numerical solutions (see Fig.~\ref{fig:mw-compare-vc}).
However, this curl-induced acceleration is almost identical for the two-field limit $m_\times \to 0$ and the single-field limit $m_\times \to \infty$.
This is why these two limits produce rotation curves that differ by much less than a percent (see Fig.~\ref{fig:mw-compare-vc}).

\begin{figure}
 \centering
 \includegraphics[width=\columnwidth]{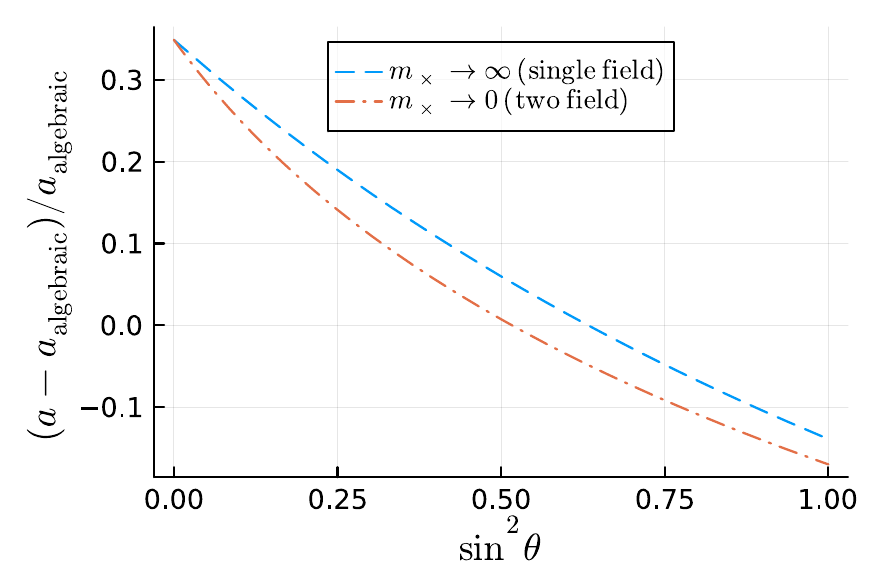}
 \caption{
   The internal acceleration in wide binaries induced by curl terms in the single-field limit $m_\times \to \infty$ (dashed blue line) and the two-field limit $m_\times \to 0$ (dash-dotted red line), assuming that the external field satisfies the algebraic MOND relation Eq.~\eqref{eq:algebraic}, i.e. assuming that only the internal acceleration is affected by curl terms.
   Here, $a$ and $a_{\mathrm{algebraic}}$ refer to the acceleration components in the radial direction connecting the two stars in the binary system.
   The difference between the curl-induced accelerations in the single-field and two-field limits is larger than for rotation curves (see Fig.~\ref{fig:mw-curl-terms}).
   This is why the value of $m_\times$ makes a larger, percent-level difference for wide binaries (see Sec.~\ref{sec:wb}).
   For $\sin^2 \theta = 0$, the curl-induced acceleration is the same in the single-field and two-field limits.
   This is why, in this case, the total internal acceleration is the same for both limits (see Fig.~\ref{fig:wb-compare-algebraic}).
 }
 \label{fig:wb-curl-terms}
\end{figure}

Consider now the internal acceleration in wide binaries.
The curl-induced part of this internal acceleration is illustrated in Fig.~\ref{fig:wb-curl-terms}.
For simplicity, Fig.~\ref{fig:wb-curl-terms} shows the case where the external field satisfies the algebraic MOND relation Eq.~\eqref{eq:algebraic}.

We see that, in wide binaries, curl terms induce a much larger change in acceleration, on the order of ten percent.
The difference in these curl terms between the single-field and two-field limits has similarly increased:
The curl terms for $m_\times \to \infty$ and $m_\times \to 0$ differ on the percent level.
This explains why we find percent-level differences between the $m_\times \to 0$ and $m_\times \to \infty$ limits for wide binaries but not for galactic rotation curves.

Fig.~\ref{fig:wb-curl-terms} also shows that, for $\sin^2 \theta = 0$, the curl terms in the $m_\times \to \infty$ and $m_\times \to 0$ limits induce exactly the same change in acceleration.
As a result, both limits give the same total acceleration in this case.
This matches our results from Sec.~\ref{sec:wb} (see Fig.~\ref{fig:wb-compare-algebraic}).

The curl terms discussed here, which explain the difference between the single-field and two-field limits, are related to -- but not identical with -- the curl part of the vector field $\vec{U}$ (or equivalently $\vec{A}$) discussed in Sec.~\ref{sec:novector} and Sec.~\ref{sec:vector}.
Indeed, in the single-field limit, the curl part of $\vec{U}$ directly matches the curl term discussed here, $\vec{U} = - f_G \, \vec{\nabla} \times \vec{h}_\infty + (\operatorname{curl-less})$.
This follows from Eq.~\eqref{eq:mudef} and Eq.~\eqref{eq:curlterm:single}.
In contrast, in the two-field limit, $\vec{U}$ is curl-less to leading order, $\vec{U} = \mathcal{O}(m_\times^2) + (\operatorname{curl-less})$ (see Eq.~\eqref{eq:Ucurl:twofield}).
But the $\mathcal{O}(m_\times^2)$ curl term in $\vec{U}$ is still related to the curl terms discussed here.
Specifically, to leading order in $m_\times^2$, we have $\vec{U} = \vec{U}_\times + (\operatorname{curl-less})$ where $\vec{U}_\times$ is determined by $\Delta \vec{U}_\times = 2 m_\times^2 \vec{\nabla} \times \vec{h}_0$.
This follows by using the Helmholtz decomposition Eq.~\eqref{eq:Udecomp} in Eq.~\eqref{eq:eom:U}, expanding in $m_\times^2$, using the fact that $\vec{U}_\times$ is of order $m_\times^2$, and comparing the result to Eq.~\eqref{eq:curlterm:two}.
Thus, the curl part of $\vec{U}$ is indeed closely related to the curl terms that explain why $m_\times$ has a larger effect in wide binaries than in galaxies.

\section{Conclusion}
\label{sec:conclusion}

Previous studies have shown that the quasi-static limit of AeST depends on a scale $m$ that is related to the so-called ghost condensate and that determines on which scales AeST reproduces MOND.
Here, we have shown that the quasi-static limit additionally depends on a scale $m_\times$ that is related to the vector field $\vec{A}$.
In previous studies, this vector field was assumed to vanish.
Above, we have argued that this assumption is, in general, not justified and we have shown how to correctly take the vector field and the associated scale $m_\times$ into account.

In the limits $m_\times \to \infty$ and $m_\times \to 0$ one recovers a single-field and a two-field version of MOND, respectively.
More precisely, these limits lead to single-field and two-field equations of motion that produce MOND-like behavior to the same extent as the equations originally derived assuming a vanishing vector field.
That is, the single-field and two-field limits still reproduce MOND as long as the ghost condensate can be neglected.
These single-field and two-field limits are relevant in systems whose characteristic length scales $l$ satisfy $m_\times l \gg 1$ and $m_\times l \ll 1$, respectively.

The two-field limit is precisely what was found in previous works where the vector field was neglected.
Thus, these previous works are justified for $m_\times l \ll 1$, but not in general.
Outside the special cases $m_\times l \gg 1$ and $m_\times l \ll 1$, the quasi-static limit of AeST represents a novel version of MOND that does not reduce to any of the previously-proposed versions such as AQUAL \citep{Bekenstein1984}, quasi-linear MOND  \citep[QUMOND,][]{Milgrom2010, Milgrom2023}, or modified inertia \citep{Milgrom2022}.

In practice, however, numerical calculations indicate that the value of $m_\times$ makes almost no difference for radial and vertical accelerations in galaxies.
Thus, for rotation curves, for example, our results are of purely theoretical interest.

In contrast, we find that there may be a percent-level difference for the acceleration in wide binaries.
This is of practical interest since observations are getting close to reaching that level of precision.
Unfortunately, for the phenomenologically relevant two-field limit $m_\times \to 0$, the acceleration in wide binaries depends sensitively on the details of the local gravitational field of the Milky Way.
Thus, quantitative theoretical predictions require numerical follow-up studies that take into account the full complexity of the solar neighborhood.

\begin{acknowledgements}
\label{sec:acknowledgements}

I thank Sabine Hossenfelder, Kyu-Hyun Chae, Stacy McGaugh, Moti Milgrom, Amel Durakovic, and Will Barker for helpful discussions.
Funded by the Deutsche Forschungsgemeinschaft (DFG, German Research Foundation) – 514562826.

\end{acknowledgements}

\appendix

\section{Numerical procedure}
\label{sec:appendix:numerics}

We numerically solve the equations of motion Eq.~\eqref{eq:eomtwo} (for the two-field limit $m_\times \to 0$) and Eq.~\eqref{eq:eomsingle} (for the single-field limit $m_\times \to \infty$) using the Julia package `Gridap.jl` \citep{Badia2020, Verdugo2022}.
We here describe our procedure for the two-field limit Eq.~\eqref{eq:eomtwo}.
The simpler single-field limit Eq.~\eqref{eq:eomsingle} works analogously.
The procedure described here also applies to the original equations Eq.~\eqref{eq:eomA0} from Ref.~\cite{Skordis2020}, since these are equivalent to those from the two-field limit Eq.~\eqref{eq:eomtwo}.

We use cylindrical coordinates $R$, $z$ that we rescale to dimensionless coordinates $R_x = R/l$, $z_x = z/l$ with the length scale $l = 10\,\mathrm{kpc}$.
We solve the equations of motion Eq.~\eqref{eq:eomtwo} in terms of the fields $\hat{\Phi} = \Phi - \alpha$ and $\alpha$.
We further rescale these fields by a factor $A = 10^{-7}$, giving $u_{\hat{\Phi}} = \hat{\Phi}/A$ and $u_\alpha = \alpha/A$.

We solve the equations in a spherical region with radius $r < 100\,\mathrm{kpc} \equiv r_{\mathrm{max}}$, i.e. $\sqrt{R_x^2+z_x^2} < 10 \equiv r_{x,\mathrm{max}}$.
We assume that solutions are symmetric under $z \to -z$ which corresponds to the $z \to -z$ symmetry of the Milky Way mass model from \cite{McGaugh2019b} that we consider here.

`Gridap.jl` expects the equations to be given in weak form.
This means we do not directly solve Eq.~\eqref{eq:eomtwo} but an integral version of these equations.
Specifically, we find functions $u_{\hat{\Phi}}$ and $u_\alpha$ for which the following integral vanishes for arbitrary test functions $v$ and $w$ that vanish at the Dirichlet boundary (i.e. at $\sqrt{R_x^2 +z_x^2} = r_{x,\mathrm{max}}$, see below),
\begin{align}
\label{eq:weakform}
 0 = \int_0^{r_{x,\mathrm{max}}} dR_x \, R_x \int_0^{\sqrt{r_{x,\mathrm{max}}^2 - R_x^2}} dz_x \, I(R_x, z_x)\,,
\end{align}
with the integrand $I(R_x, z_x)$ given by
\begin{align}
\begin{split}
 I(R_x, z_x) \equiv
 &- (\partial_i v) \cdot (\partial_i u_{\hat{\Phi}}) \\
 &- (\partial_i w) \cdot \left( \tilde{\mu}\left(\frac{A |\vec{\nabla} u_\alpha|}{l a_0}\right) \partial_i u_\alpha \right) \\
 &- (v+w) \cdot \frac{4 \pi G_N l^2}{A} f_G \, \rho_b(R_x, z_x)
  \,.
\end{split}
\end{align}
Here, the subscript $i$ of $\partial_i$ runs over $R_x$ and $z_x$.
Similarly, $|\vec{\nabla} u_\alpha|$ means $\sqrt{(\partial_i u_\alpha) (\partial_i u_\alpha)}$.
We left out the condensate density $\rho_c$ since we are only interested in the case $m^2/f_G = 0$.
The baryonic energy density $\rho_b(R,z)$ is that of the Milky Way model from \cite{McGaugh2019b}.
We show how this integral form of Eq.~\eqref{eq:eomtwo} is derived below.

The integral in Eq.~\eqref{eq:weakform} is evaluated on a mesh that we generate using `Gmsh` \citep{Geuzaine2009}.
The mesh has size $R_{x,\mathrm{max}}$ and is generated with a `mesh\_size\_callback` function that returns the default cell size or $0.005 \cdot (1 + \sqrt{R_x^2 + z_x^2})$, whichever is smaller.

On the boundary $\sqrt{R_x^2+z_x^2} = r_{x,\mathrm{max}}$, we impose a constant value for $u_\alpha$ and $u_{\hat{\Phi}}$.
That is, we impose that the fields are spherically symmetric there.
Which values we impose does not matter because the absolute values of $u_\alpha$ and $u_{\hat{\Phi}}$ are inconsequential in the $m^2/f_G = 0$ case we consider here.
We choose $u_\alpha = 4.1$ and $u_{\hat{\Phi}} = -5.0$.
On the other boundary, $z_x = 0$, the form of the integral Eq.~\eqref{eq:weakform} implicitly imposes homogeneous Neumann boundary conditions which encode the $z \to -z$ symmetry (see below).

We now discuss how to obtain the integral form Eq.~\eqref{eq:weakform} of the equations of motion Eq.~\eqref{eq:eomtwo}.
We first multiply the two equations of motion in Eq.~\eqref{eq:eomtwo}, respectively, by two arbitrary test functions $v$ and $w$.
Then, we add up these two equations and integrate both sides of the resulting equation over the volume $V$ in which we want to solve the equations.
This gives
\begin{align}
\label{eq:weakformderivation1}
\begin{split}
 0 &= \int_V dV \Big\{
  v \cdot \left[
    \Delta \hat{\Phi} - 4 \pi G_N f_G (\rho_b + \rho_c)
   \right]\\
 &+ w \cdot \left[
  \vec{\nabla} \left( \tilde{\mu}\left(\frac{|\vec{\nabla} \alpha|}{a_0}\right) \vec{\nabla} \alpha \right) - 4 \pi G_N f_G (\rho_b + \rho_c)
   \right]
   \Bigg\} \,.
\end{split}
\end{align}
Finding fields $\hat{\Phi}$ and $\alpha$ that solve the two equations of motion Eq.~\eqref{eq:eomtwo} is equivalent to finding fields $\hat{\Phi}$ and $\alpha$ for which the integral Eq.~\eqref{eq:weakformderivation1} is zero for arbitrary test functions $v$ and $w$.
From Eq.~\eqref{eq:weakformderivation1} one finds the integral form Eq.~\eqref{eq:weakform} used in our numerical procedure by integrating by parts, neglecting $\rho_c$, using cylindrical coordinates, assuming axisymmetry, and switching to the rescaled fields $u_{\hat{\Phi}}$ and $u_\alpha$ and the rescaled coordinates $R_x$ and $z_x$.

The only subtlety lies in the boundary term that one obtains when integrating by parts.
This boundary term is proportional to
\begin{align}
 \label{eq:weakformboundary}
 \int_{\partial V} d\vec{S} \cdot \left[ v \, \vec{\nabla} \hat{\Phi} + w \, \tilde{\mu}\left(\frac{|\vec{\nabla} \alpha|}{a_0}\right) \vec{\nabla} \alpha \right] \,.
\end{align}
How to handle this boundary terms depends on the type of boundary conditions one wants to impose and the precise mathematical functional spaces from which one chooses the test functions $v$ and $w$ and the fields $\hat{\Phi}$ and $\alpha$.
In our case, since we impose mixed Dirichlet and homogeneous Neumann boundary conditions, the correct procedure is to leave out the boundary term Eq.~\eqref{eq:weakformboundary}, use test functions $v$ and $w$ that vanish on the Dirichlet boundary (i.e. at $r = r_{\mathrm{max}}$), and consider fields $\hat{\Phi}$ and $\alpha$ from a functional space that only contains fields that satisfy the Dirichlet boundary condition.
Unfortunately, there are some subtleties in rigorously deriving this result.
For these, we refer to the mathematical literature, see for example Ref.~\cite{Salsa2022}.
In practice, one can follow the nice tutorial\footnote{
  \url{https://gridap.github.io/Tutorials/stable/pages/t004_p_laplacian/}
}
provided by `Gridap.jl` \citep{Badia2020, Verdugo2022} on precisely the type of equation we are dealing with here.

We have validated this numerical procedure against a previous numerical code used in Ref.~\cite{Hossenfelder2020} where a very similar set of equations was solved numerically.
The numerical code from Ref.~\cite{Hossenfelder2020} does not require deriving the integral form of these equations.
Thus, that both numerical codes agree indicates that Eq.~\eqref{eq:weakform} is indeed the correct integral form of the equations of motion Eq.~\eqref{eq:eomtwo}.
The advantage of the numerical procedure used here (based on `Gridap.jl` and Eq.~\eqref{eq:weakform}) is that it runs faster and is more flexible.

\section{More general external fields}
\label{sec:appendix:generalefe}

In Sec.~\ref{sec:wb}, we assumed the same limit -- $m_\times \to 0$ or $m_\times \to \infty$ -- for both the internal and the external fields of wide binary systems.
But, in principle, it can happen that the external field cannot be calculated using the same limit as the internal field (see Sec.~\ref{sec:scale}).
Here, we show that in such a case Eq.~\eqref{eq:aoveraNsingle} and Eq.~\eqref{eq:aoveraNtwo} remain valid.
Only the external fields $x_e$ and $x_{\varphi,e}$ that enter these equations need to be calculated differently.

To see this, we first write down the equations of motion Eq.~\eqref{eq:eom} without assuming a particular value of $m_\times$ for a system embedded in external fields $\vec{U}_e$ and $\Phi_e$, giving
\begin{align}
 \Delta \Phi - \vec{\nabla} \vec{U} &= 4 \pi G_N f_G (\rho_b + \rho_c) \,,
\end{align}
and
\begin{multline}
\vec{\nabla} \Phi - \frac{1}{2 m_\times^2} \vec{\nabla} \times \vec{\nabla} \times \vec{U} =\\
\vec{U}  + (\vec{U} + \vec{U}_e) \, \tilde{\mu}\left(\frac{|\vec{U} + \vec{U}_e|}{a_0}\right) - \vec{U}_e \, \tilde{\mu} \left(\frac{|\vec{U}_e|}{a_0}\right) \,,
\end{multline}
where we used that $\rho_c$ is linear in the field $\Phi$.
As in Sec.~\ref{sec:wb}, we assume that the external field dominates everywhere.
For this case, we find the linear equations
\begin{multline}
\label{eq:eomefdomgeneral}
\begin{aligned}
 \Delta \Phi - \vec{\nabla} \vec{U} &= S\,, \\
 \vec{\nabla} \Phi - \frac{1}{2 m_\times^2} \vec{\nabla} \times \vec{\nabla} \times \vec{U} &=%
 \end{aligned}%
 \\[-1em]%
  \vec{U}  + \vec{U} \, \tilde{\mu}_e + (\vec{U} \cdot \vec{U}_e) \frac{\hat{U}_e}{a_0} \tilde{\mu}'_e \,,
\end{multline}
where $\hat{U}_e$ denotes the direction of $\vec{U}_e$ and we use the shorthand notations
\begin{align}
 \tilde{\mu}_e \equiv \tilde{\mu}\left(\frac{|\vec{U}_e|}{a_0}\right) \,, \quad
 \tilde{\mu}_e' \equiv \tilde{\mu}'\left(\frac{|\vec{U}_e|}{a_0}\right) \,,
\end{align}
as well as
\begin{align}
 S \equiv 4 \pi G_N f_G (\rho_b + \rho_c) \,.
\end{align}

We now consider wide  binary systems with typical length scale $l_{\mathrm{wb}}$ and consider the cases where $m_\times l_{\mathrm{wb}}$ is large or small without assuming anything about the external field.
Below, we show that in this case we still reproduce Eq.~\eqref{eq:aoveraNsingle} for $m_\times l_{\mathrm{wb}} \gg 1$ and Eq.~\eqref{eq:aoveraNtwo} for $m_\times l_{\mathrm{wb}} \ll 1$, but with adjusted definitions of $x_\Phi$ and $x_{\varphi,e}$ compared to Sec.~\ref{sec:wb}.
For concreteness, we assume that $\vec{U}_e$ points in the positive $z$ direction.

\subsection{The case $m_\times l_{\mathrm{wb}} \ll 1$}

Consider first the case $m_\times l_{\mathrm{wb}} \ll 1$.
Following the same steps as in Sec.~\ref{sec:twofield}, we find from Eq.~\eqref{eq:eomefdomgeneral} that $\vec{U} = \vec{\nabla} \alpha + \mathcal{O}((m_\times l_{\mathrm{wb}})^2)$, which leads to
\begin{subequations}
\begin{align}
\Delta (\Phi - \alpha) &= S \,, \\
\Delta (\Phi - \alpha) &= \tilde{\mu}_e (\partial_x^2 \alpha + \partial_y^2 \alpha) +  \tilde{\mu}_e (1 + \tilde{L}_e ) \partial_z^2 \alpha \,,
\end{align}
\end{subequations}
with the shorthand notation
\begin{align}
 \tilde{L}_e = \frac{|\vec{U}_e|}{a_0} \frac{\tilde{\mu}'_e}{\tilde{\mu}_e} \,.
\end{align}
This is equivalent to
\begin{subequations}
\begin{align}
\Delta (\Phi - \alpha) &= S \,, \\
\tilde{\mu}_e (\partial_x^2 \alpha + \partial_y^2 \alpha) +  \tilde{\mu}_e (1 + \tilde{L}_e ) \partial_z^2 \alpha &= S \,.
\end{align}
\end{subequations}
Following the procedure of \cite{Banik2015} and neglecting the ghost condensate density $\rho_c$, this gives the two-field expression for $a/a_N$ from Eq.~\eqref{eq:aoveraNtwo}, just with the replacement
\begin{align}
 x_{\varphi,e} \to \frac{|\vec{U}_e|}{a_0} \,.
\end{align}

\subsection{The case $m_\times l_{\mathrm{wb}} \gg 1$}

Consider next the case $m_\times l_{\mathrm{wb}} \gg 1$.
In this case, we find from Eq.~\eqref{eq:eomefdomgeneral},
\begin{subequations}
\begin{align}
\Delta \Phi - \vec{\nabla} \vec{U} &= S \,, \\
\partial_k \Phi &= U_k (1 + \tilde{\mu}_e) \,, \\
\partial_z \Phi &= U_z (1 + \tilde{\mu}_e) + U_z \tilde{\mu}_e \tilde{L}_e \,,
\end{align}
\end{subequations}
where $k = x,y$.
We can algebraically solve for $\vec{U}$ and plug the result back into the first of these equations to find
\begin{multline}
\left(1 - \frac{1}{1 + \tilde{\mu}_e}\right) (\partial_x^2 \Phi + \partial_y^2 \Phi) + \\
  \left(1 - \frac{1}{1 + \tilde{\mu}_e(1 + \tilde{L}_e)}\right) \partial_z^2 \Phi = S \,.
\end{multline}
To make contact with Eq.~\eqref{eq:aoveraNsingle}, we must rewrite all occurrences of $\tilde{\mu}$ in terms of $\mu$.
In Sec.~\ref{sec:singlefield}, the function $\mu$ is implicitly defined in terms of the function $\tilde{\mu}$ by the relations
\begin{subequations}
\begin{align}
 x &\equiv u \left( 1 + \tilde{\mu}(u) \right) \,, \\
 u &= x \left(1 - f_G \mu(x) \right) \,,
\end{align}
\end{subequations}
for all positive $u$.
From both relations we can infer an expression for $u/x$.
Equating these gives a useful direct relation between $\tilde{\mu}$ and $\mu$,
\begin{align}
 (1 + \tilde{\mu}(u))(1  - f_G \mu(x)) = 1 \,.
\end{align}
By taking a derivative of this relation with respect to $x$ and using $\partial u / \partial x = 1 - f_G \mu(x) (1 + L(x))$ we find
\begin{align}
 1 + \tilde{L}(u) = (1 + L(x)) \frac{1 - f_G \mu(x)}{1 - f_G \mu(x) (1 + L(x))} \,.
\end{align}
After some algebra, we then obtain the relations
\begin{align}
\begin{split}
 1 - \frac{1}{1 + \tilde{\mu}(u)} &= f_G \mu(x) \,, \\
 1 - \frac{1}{1 + \tilde{\mu}(u)(1 + \tilde{L}(u))} &= f_G \mu(x) (1 + L(x)) \,.
\end{split}
\end{align}
Thus, we have
\begin{align}
\mu_e (\partial_x^2 \Phi + \partial_y^2 \Phi) + \mu_e (1 + L_e) \partial_z^2 \Phi = S \,.
\end{align}
with the shorthand notation
\begin{align}
\begin{split}
 \mu_e &\equiv \mu(x_u) \,, \\
 L_e &\equiv \frac{x_u \mu'(x_u)}{\mu(x_u)} \,, \\
 x_u &\equiv \frac{|\vec{U}_e|}{a_0} \left( 1 + \tilde{\mu}\left(\frac{|\vec{U}_e|}{a_0}\right)\right) \,.
\end{split}
\end{align}
Following the procedure of \cite{Banik2015} and neglecting the ghost condensate density $\rho_c$, this gives the single-field expression for $a/a_N$ from Eq.~\eqref{eq:aoveraNsingle}, just with the replacement
\begin{align}
 x_e \to \frac{|\vec{U}_e|}{a_0} \left( 1 + \tilde{\mu}\left(\frac{|\vec{U}_e|}{a_0}\right)\right)\,.
\end{align}

\bibliography{aest-static}

\end{document}